\documentclass[11pt]{article}
\usepackage[margin=1.0in]{geometry} 
\usepackage{ytableau}
\usepackage{color}
\usepackage{fullpage}
\usepackage{hyperref} 
\definecolor{darkred}{rgb}{0.8,0,0}
\definecolor{darkgreen}{rgb}{.0,.8,.0}
\hypersetup{ colorlinks=true, linkcolor=blue, citecolor=red } 
\usepackage{amsmath, amsthm, amssymb,amsfonts}
\usepackage{epsfig}
\usepackage{graphicx}

\newcommand{\hf}{\frac{1}{2}}

\newcommand{\nb}{{\bar {n}}}
\newcommand{\mb}{{\bar {m}}}
\newcommand{\xn}{x_{n}}
\newcommand{\xnb}{{\bar x _{n}}}

\newcommand{\xm}{x_{m}}

\newcommand{\e}{e^{i k_{0}Y}}                                      

\newcommand{\kim}{ k_{1\mu}}                                      
\newcommand{\kom}{ k_{0\mu}}                                      
\newcommand{\ki}{ k_{1}}
\newcommand{\yi}{ Y_{1}}
\newcommand{\yib}{ Y_{\bar 1}}

\newcommand{\kn}{ k_{n}}

\newcommand{\kt}{ k_{2}}                                             
\newcommand{\ko}{ k_{0}}                                             
\newcommand{\yim}{ Y_{1}^{\mu}}                                      
\newcommand{\yin}{ Y_{1}^{\nu}}                                      
\newcommand{\kin}{ k_{1\nu}}  
\newcommand{\kir}{ k_{1\rho}} 
                                    
\newcommand{\kon}{ k_{0\nu}}
\newcommand{\kor}{ k_{0\rho}}                                      
\newcommand{\ktm}{ k_{2\mu}}   
 \newcommand{\ktn}{ k_{2\nu}}

\newcommand{\lpp}{\mbox {$e^{i\int _{c} \alpha (t)                             
k(t) \partial _{z} X(z+t) dt +ik_{0}X}$}}

\newcommand{\gvk}{ e^{i\sum _{n }k_{n}Y_{n}}}

\newcommand{\p}{\partial}                                           
\newcommand{\pp}{\partial ^{2}}

\newcommand{\li}{ \lambda_{1}} 
\newcommand{\lib}{ \lambda_{\bar 1}}                                    
\newcommand{\lt}{ \lambda_{2}} 
\newcommand{\ltb}{ \lambda_{\bar 2}}                                    
\newcommand{\eps}{ \epsilon}                                        
\newcommand{\al}{\alpha }                                             
\newcommand{\aln}{\alpha _{n}} 
                                 
\newcommand{\lan}{\langle}
\newcommand{\ran}{\rangle}

\newcommand{\xb}{\mbox{$\bar{x}$}}

\newcommand{\zb}{{\bar{z}}}                                             
\newcommand{\tb}{\mbox{$\bar{t}$}}

\newcommand{\kimb}{\mbox {$ {k_{\bar1\mu}}$}}                               
\newcommand{\kinb}{\mbox {$ {k_{\bar1\nu}}$}}  
\newcommand{\kirb}{\mbox {$ {k_{\bar1\rho}}$}}                               
\newcommand{\kisb}{\mbox {$ {k_{\bar1\sigma}}$}}

\newcommand{\kib}{\mbox {$ {k_{\bar1}}$}}                                      
  
\newcommand{\ktrb}{\mbox {$ {k_{\bar 2\rho}}$}}                                      
\newcommand{\qt}{ q_{2}}                                             
\newcommand{\qi}{\mbox {$ q_{1}$}}                                             
\newcommand{\qtb}{\mbox {$ \bar{q_{2}}$}}                                      
\newcommand{\qib}{\mbox {$ \bar{q_{1}}$}}                                      
\newcommand{\qo}{ q_{0}}

\newcommand{\la}{ \lambda }                                           
\newcommand{\be}{\begin{equation}}                                             
\newcommand{\br}{\begin{eqnarray}}                                             
\newcommand{\ee}{\end{equation}}                                               
\newcommand{\er}{\end{eqnarray}}

\begin{document}
\title{
\hfill\parbox{4cm}{\normalsize IMSC/2014/07/05\\
}\\
\vspace{2cm}
Gauge Invariance and Equations of Motion for Closed String Modes
}
\author{B. Sathiapalan\\ {\em                                                  
Institute of Mathematical Sciences}\\{\em Taramani                     
}\\{\em Chennai, India 600113}}                                     
\maketitle                                                                     
\begin{abstract}   
We continue earlier discussions on loop variables and the exact renormalization group on the string world sheet for closed and open string backgrounds.  The world sheet action with a UV regulator is written in a generally background covariant way by introducing a background metric. It is shown that the renormalization group gives background covariant equations of motion - this is the gauge invariance of the graviton. Interaction is written in terms of gauge invariant and generally covariant field strength tensors. The basic idea is to work in Riemann normal coordinates and covariantize the final equation. It turns out that the equations for massive modes are gauge invariant only if the space time curvature of the (arbitrary) background is zero. The exact RG equations give  quadratic equations of motion for all the  modes {\em including} the physical graviton. The level $(2,\bar 2)$ massive field equations are  used to illustrate the techniques.  At this level there are  mixed symmetry tensors. Gauge invariant interacting equations can be written down. In flat space an action can also be written for the free theory.

 \end{abstract}  
 
\newpage 
\tableofcontents 
                                                             
\newpage                                                                       
\section{Introduction} 

The Renormalization Group (RG) approach to obtaining equations of motion for the fields of string theory has a long history [\cite{L}-\cite{T}]. Further developing these ideas the Exact Renormalization Group \cite{WK,W,W2,P} has also been used fruitfully \cite{BM,HLP}. The connection between $\beta$ functions, the equations of motion and Zamolodchikov metric was described in \cite{Poly} and also shown in the case of constant gauge fields in \cite{ACNY}.  Using similar ideas a proof that the tachyon equation of motion is given by the beta function multiplied by the Zamolodchikov metric was given in \cite{BSPT}. 

The issue of gauge invariance in string theory was dealt with comprehensively using string field theory in \cite{SZ,WS,WS2,WS3,WS4,Wi,Wi2,BZ} via the BRST formalism. An action has been written down for open strings \cite{Wi2} and also closed strings \cite{BZ}.

One can ask whether a manifestly background independent formalism - in the same sense that Einstein's equation is background independent - exists. This means that it (the formalism) is not tied down to any particular starting background. In particular the background need not be a solution to the equations of motion. So it should not be necessary to start with a background that is a 2D CFT on the world sheet. Background independence in the BRST formalism has been discussed in \cite{Wi3,LW,Sh}. One of the issues here that needs to resolved is the clash between BRST invariance and dependence of the world sheet theory on a cutoff, when the fields are off shell.

The RG approach is manifestly background independent. Using the loop variable approach it was also shown to be gauge invariant \cite{BSLV}
at the free level. More recently  in \cite{BSERGopen1,BSERGopen2,BSERGclosed} (hereafter I,II,III) the loop variable formalism was used to construct gauge invariant interacting EOM using the exact renormalization group (ERG). The salient features of this formalism are as follows:
\begin{enumerate}
\item
The fields for the open strings are described in terms of loop variables $k_\mu(t)= \kom +{\kim \over t} +...+{k_{n\mu}\over t^n} +.. $ and have the invariance $k_\mu (t) \rightarrow \la (t) k_\mu (t), \la(t) = 1 + {\li\over t} + {\la_n\over t^n}+... $. For closed strings there is also the anti holomorphic part $\bar k_\mu(\bar t)= \kom +{\bar k_{1 \mu} \over \bar t} +...+{\bar k_{n\mu}\over {\bar t}^n} +.. $ and corresponding gauge invariance parametrized by $\bar \la (\bar t)$. $\kom$ is the usual momentum.

\item
The coordinate $X^\mu(z)$ is generalized to a coordinate $Y^\mu(z,x_1,...,\xn,..)$ which obeys ${\pp Y\over \p \xn \p \xm}= {\p Y\over \p x_{n+m}}$. For closed strings one has $Y(z,\bar z, \xn , \bar x _n)$.

\item
The free level equations look massless in one extra dimension. To obtain the string spectrum, one has to dimensionally reduce. Thus $k^\mu (t), \mu=1...D+1$ becomes $k^\mu(t), q(t), \mu=1...D$. $q(t)= q_0+{q_1\over t} +{q_2\over t^2}+...+{q_n\over t^n}+...$. And
 $q_0$ is the mass. The gauge transformation of $q(t)$ is $q(t)\rightarrow \la(t)q(t)$.

\item To match with the BRST field content one has to get rid of $q_1$ in a way that is consistent with gauge invariance. These rules were worked out up to level 5. The fact that it can be done in a way that is consistent with gauge invariance turns out to be non trivial in that it required that an overdetermined set of linear equations have a consistent solution. It was also not obvious a priori that this solution should turn out to be consistent with the statement that the theory is obtained by dimensional reduction from one higher dimension. Surprisingly it turns out to be consistent. This seems to point to a higher dimensional origin for the theory. 

\item It has been shown earlier \cite{BSOC} that if one wants to match the constraints and gauge transformations with those of string theory, one obtains $D=26$ and $q_0^2=2,4$ for the first two massive levels of the open string.  
 
\item
The ERG is quadratic, hence the interacting equations of motion (EOM) are also quadratic. This is true for closed strings also.

\item
A significant feature in both the open and closed string cases is that gauge invariance of the interacting theory, i.e. gauge invariance of the "field strength", requires that one introduce extra loop variables $K_{n_1n_2...n_k \mu}$ dual to $\p^k Y^\mu\over \p x_{n_1} ...\p x_{n_k}$ with specific gauge transformations. (In the closed string case we also have loop variables parametrized by both holomorphic and anti-holomorphic indices corresponding to $\xn$ and $\bar x_n$.) The construction of these variables are possible only when $q_0$, the mass, is non zero. Thus dimensional reduction with mass is forced on us by the interactions. 

\item
For closed strings, at the intermediate stages of the calculation, one also has mixed derivatives corresponding to vertex operators of the form $\pp X\over \p z \p \bar z$. This is because at the intermediate stages the regulator breaks the factorization into holomorphic and anti-holomorphic parts. This can be seen in the regulated two point correlator $\langle X(z,\bar z) X(0)\rangle \sim ln~(z\bar z + a^2)$.  These states do not contribute to the S-matrix of physical states.

\item
For open strings the gauge transformations are those of the free theory and are not modified by interactions, unlike Witten's BRST string field theory. In this sense the theory looks Abelian, until Chan-Paton factors are introduced. The interactions are written in terms of gauge invariant "field strengths".

\item For closed strings, the gauge transformations need to be modified to include a "non-Abelian rotation". This is the usual transformation induced on tensor fields by general coordinate transformations. The connection between "gauge" symmetries and general coordinate invariance of general relativity becomes evident for the first time. However it turns out that a background reference metric has to be introduced and the manifest symmetry is not the usual general coordinate invariance but a {\em background} general coordinate invariance. One has to use the Riemann normal coordinate expansion \cite{Pet,Eis,AGFM} to make sure that the equations are background covariant.  Thus the equations are non polynomial in the background metric, but is at most quadratic in the physical graviton field. 

\item  It was found that non zero curvature of the background metric spoils gauge invariance of the massive modes. Since the background metric curvature is unrelated to the physical metric curvature it was convenient to set it to zero. Due to this limitation the formalism was not background independent. 

\item
In III, it was argued that the full two dimensional world sheet action is independent of this background metric in the continuum limit. So  the original world sheet action has the full general coordinate invariance (in addition to background general coordinate invariance). Thus we expect that the low energy EOM for the graviton which is non polynomial, will have the full covariance. At the intermediate stages of the calculation, it will have only the background covariance.

\end{enumerate}
In this paper we continue the discussion. The main purpose is to clarify the details for the closed string massive mode equations, which were only sketched out in the earlier paper and work out some examples. The equations have to be background covariant. At the same time the action should not have a dependence on the background metric. This turns out to be straightforward if curvature of the background is zero.\footnote{There is no apriori connection between general coordinate invariance and  curvature of space-time. One is a gauge symmetry of the theory and equations of motion. The other is a physical (or geometrical) property of a solution of the theory or of  a background configuration used often as a first approximation to a solution.} 
The ERG  necessarily involves a regulator and one has to show that this is consistent with general coordinate invariance. Otherwise there could be anomalies and the final answer will not be general coordinate invariant and we will not recover Einstein's equation for the graviton. (This is point 11 above.) We show that in fact the regulated action can be made general coordinate invariant.  The theory is regulated by  adding higher derivative terms to the kinetic term. This modifies the short distance behaviour of the two point function and regulates the theory. It is shown that this corresponds to some background values for massive fields. The coordinate invariance of the theory can be maintained, if one modifies the transformations laws for the massive fields to include additional non tensorial terms. Alternatively a modified
field has to be defined that has tensorial properties.

The crucial requirement for gauge invariance of the theory is that the derivatives $\p\over \p \xn$ that obey ${\pp Y\over \p \xn \xm}={\p Y \over \p x_{n+m}}$ need to be covariantised while retaining this property. As was shown in III this is also easy to achieve if the background curvature is zero. In practice, once we know that the action has general covariance, we can work in Riemann Normal Coordinates (RNC).
All derivatives can be interpreted as covariant derivatives and then the equations are valid in any coordinate system. The only subtlety is that in the interaction term we have fields at two different points and one must do a covariant Taylor expansion. This will be discussed in detail. Thus we will show in some detail that in the case of flat background metric the massive mode equations will be both gauge and general covariant.  

In this case the metric perturbation is necessarily about a flat background and hence not background independent. The interesting thing is that the physical graviton  occurs only quadratically just as any other massive mode.

As an application of this formalism, we work out in some detail the equations of motion for the level ($2,\bar 2$) fields. The details of the dimensional reduction, field content and gauge transformation are also worked out. Since this level involves fields of mixed symmetry, this is interesting quite independent of string theory. It turns out that starting from the free equation of motion it is not too hard to write down an action for the mixed symmetry field.

This paper is organized as follows: In Section 2 we give a recapitulation of some material in the earlier papers. In Section 3 we outline the five steps involved.  In Sec 4 we discuss Step 1 which is implementing general covariance. In Section 5 we demonstrate how to regulate the theory, consistent with general coordinate invariance - this is Step 2. In Section 6 we discuss Step 3 which is the modification of the transformation laws for fields. In Section 7  we work out the ERG and details of the covariant Operator Product expansion needed for the interaction term.   Section 8 contains an application of this procedure to a specific case: we give details of the field content and interacting equations for level 4 fields of the closed string. Section 9 contains some conclusions and open questions.

\section{Recapitulation}
\subsection{Exact Renormalization Group Equation}
The equation is written in position space. We let $z = x$ for open strings and $z=x+iy$ for closed strings. Thus $\int dz$ is to be understood as $\int d^2z$ for closed strings. Also $X(z)=X(z,\bar z)$. 

 The action is:
\[
S= \underbrace{-\hf \int dz ~\int dz' ~Y^\mu(z) (G^{-1})_{\mu\nu}(z,z';\tau)Y^\nu(z')}_{Kinetic~ term=K} + \underbrace{\int dz~L[Y^\mu(z),Y^\mu_{n,\bar m}(z)]}_{Interaction=S_{int}}
\]
$G^{\mu\nu}(z,z';\tau)\equiv\lan Y^\mu(z) Y^\nu(z')\ran$ is a cutoff propagator, where $\tau$ parametrizes the cutoff (eg $\tau = ln ~a$). We let $\mu =0,...,D-1$ be the usual space time coordinate and when $\mu=D$, let $Y^D=\theta$ be the extra dimension. $\theta$ is assumed to be massive world sheet field so that it's Green function $G^{DD}(z,z';\tau)=\langle \theta(z)\theta(z')\rangle$ falls off exponentially on a scale of the world sheet cutoff $a$. This is important: We want the low energy theory on the world sheet to be that
of 26 massless scalars in order to reproduce the Veneziano amplitude and its generalizations. The Green function will however contribute in self contractions within a vertex operator and will enter in the mass shell and physical state constraints.

The ERG is the following:
\[
\int du ~ {\p L[X(u)]\over \p \tau} =
\]
\be	\label{ERG}
 \int dz~\int dz'~\hf~\dot G^{\mu \nu}(z,z')\Bigg( \int du~{\delta^2L[X(u)] \over \delta X^\nu(z')\delta X^\mu(z)}+
\int du~\int dv~ {\delta L[X(u)]\over \delta X^\mu(z)}{\delta L[X(v)]\over \delta X^\nu(z')}\Bigg)
\ee

 where $\dot G^{\mu\nu} \equiv {\p G^{\mu\nu}\over \p \tau}$.
 
 \subsection{Loop Variables for Open Strings}

The following equations summarize the facts about loop variables for open strings: 
\be   \label{LV}
e^{i{\cal L}[X(z)]}=\lpp
\ee
with
\be	\label{k}
k(t) = \ko + {\ki \over t} + {\kt \over t^2} +....+{\kn\over t^n}+...
\ee
and 
\be	\label{al}
\al(t)=e^{\sum x_n t^{-n}}\equiv 1+ {\al_1\over t}+...+{\aln\over t^n}+...
\ee
$\aln$ satisfy:
${\p \aln\over \p x_p}= \al_{n-p}$.

\be	\label{Y}
Y\equiv X(z) + \al _1 \p_zX(z) + \al _2  \p_z^2 X(z) + {\al _3 \p_z^3X(z)\over 2!}+...+{\aln \p_z^n X(z)\over (n-1)!}+...
\ee

with $Y_n= {\p Y\over \p \xn}$. 
Thus 
\be
\lpp = \gvk
\ee
$Y$ has the crucial property that ${\pp Y\over \p \xn \p \xm}= {\p Y\over \p x_{n+m}}$.

However in I we had also introduced  $Y_{n_1,n_2}= {\pp Y\over \p _{x_{n_1}}\p_{x_{n_2}}}$ and so on. It turns out that introducing these
separately is important for gauge invariance of the interaction term. Dual to these vertex operators are $K_{n_1n_2...}$ which are linear combinations of the $\kn$ and $q_n$. The exact expressions are given in I and II and we will use them in Section 5.

Thus for open string loop variables  $z\equiv \{z, \xn\}$. 
\be  \label{zxn}
\int dz \equiv \int dz~ \prod _{n=1,2..}\int d\xn
\ee
Also
\be
{\delta \over \delta Y(z)} Y(z') = \delta (z-z')
\ee
where
\be
\delta (z-z')\equiv \delta (z-z')\prod _{n=1,2..}\delta(x_n-x'_n)
\ee

Keeping these substitutions in mind one can use the ERG for open string loop variables.

\subsection{Loop Variables for Closed Strings}

For closed strings we introduce the anti holomorphic variables and write:
\be	
\int dz \equiv \int d^2z~ \prod _{n=1,2..}\int d\xn \int d \bar x_n
\ee
Also
\be
{\delta \over \delta Y(z)} Y(z') = \delta (z-z')
\ee
where
\be	\label{zxnb}
\delta (z-z')\equiv \delta^2 (z-z')\prod _{n=1,2..}\delta(x_n-x'_n)\delta(\bar x_n- \bar x'_n)
\ee
Using these definitions, the ERG can be written down for any given functional.

Apply this to functional  $\int du~L[Y(u),Y_{n,\mb}(u)]$: ($\xn$ will be associated with $u$, $\xn'$ with $z'$ and $\xn ''$ with $z''$)
\[
{\delta \over \delta Y(z')} \int du~ L[Y(u),Y_{n;\mb}(u)]=\]
\[
\int du~\Big\{ {\p L [Y(u),Y_{n;\mb}(u)]\over \p Y(u)} \delta (u-z')+\]
\[
\sum_ {n=1,2,...} {\p L [Y(u),Y_{n;\mb}(u)]\over \p Y_n(u)} \p_{\xn}\delta (u-z')+
\sum_ {n_1,n_2=1,2,...} {\p L [Y(u),Y_{n;\mb}(u)]\over \p Y_{n_1,n_2}(u)} \p_{x_{n_1}}\p _{x_{n_2}}\delta (u-z')\]
\[
+\sum_ {\mb =1,2,...}
 {\p L [Y(u),Y_{n;\mb}(u)] \over \p Y_{\nb} (u) }
 \p_{\bar x _m}\delta (u-z') 
  +\sum_ {\mb_1,\mb _2=1,2,...}
 {\p L [Y(u),Y_{n;\mb}(u)] \over \p Y_{\mb_1,\mb_2} (u) }
 \p _{\bar x_{m_1}}\p_{\bar x_{m_2}} \delta (u-z')
+\]\be    \label{FD}
\sum_ {n,\mb=1,2,...}
 {\p L [Y(u),Y_{n,\mb}(u)] \over \p Y_{n,\mb} (u) }
 \p _{ x_{n}}\p_{\bar x_{m}} \delta (u-z') +...\Big\}\ee

As mentioned in the introduction, we have kept open the possibility of mixed holomorphic-anti-holomorphic derivatives, since these are in fact needed for gauge invariance of the closed string equations.

The loop variable turns out to be a generalization of the open string one (We have suppressed the Lorentz index $\mu$ below):
\[
e^{i{\cal L}[X(z)]}=Exp ~\Big(i\Big(\ko . X (z) + \oint_c dt~ k(t) \al(t) \p_z X(z+t) + \oint_c d\bar t~ \bar k(\bar t) \bar \al(\bar t) \p_\zb X(\zb+\bar t)+ 
\]
\be	\label{LVC}
+\oint_c dt\oint_c d\bar t~K(t,\bar t) \al(t)
\bar \al (\bar t) \p_z \p_\zb X(z+t,\zb+\tb)\Big)\Big)
\ee
where
\be	\label{K}
K(t,\tb)\equiv K_{0;0}+\sum _{\mb=1}^\infty K_{0;\mb}\tb^{-\bar m} + \sum _{n=1}^\infty K_{n;0}t^{-n} + \sum _{n=1,\mb=1}^\infty K_{n;\mb}t^{-n}\tb^{-\mb}
\ee
and $\bar\al (\bar t)$ is the anti-holomorphic counterpart of $\al(t)$ defined for open strings.
$K_{n;0}$ and $K_{0;\mb}$ are the $\kn$ and $\bar k_{\bar m}$ mentioned in the introduction.
If we define 
\be \label{Y}
Y=
\Big(X+ \al_1 \p_zX + \al_2 \p_z^2 X +{\al_3\p_z^3X\over 2!}+...+\bar \al_1 \p_zX + \bar \al_2 \p_\zb^2X+...+{\al_n\bar \al_m\p_z^n\p_\zb^mX\over (n-1)!(m-1)!}+..\Big)
\ee
we can write the closed string loop variable as:
\be  \label{GVC}
Exp\Big(i\Big( \ko.Y + K_{1;0}.{\p Y\over \p x_1}+K_{0;\bar 1}.{\p Y\over \p \bar x_1}+K_{1;\bar 1}.{\pp Y\over \p x_1\p \bar x_1}+...
+K_{n;\mb}.{\p^2Y\over \p \xn \p\bar x_m}+...\Big)\Big)
\ee

While ${\p^4 Y\over \p x_{n_1} \p x_{n_2}\p \xb _{m_1}\p \xb_{m_2}}={\pp Y\over \p x_{n_1+n_2}\p \bar x_{m_1+m_2}}$ continues to be true here as for open strings, again just as for open strings we will need to separately introduce  vertex operators 
${\p\over \p x_{n_1}}{\p\over \p x_{n_2}}...{\p\over \p \bar x_{m_1}}{\p \over \p \bar x_{m_2}} ...Y$ and $K_{n_1,n_2,...;\bar m_1,\bar m_2,...}$ as their coefficients. Expressions for $K_{[n]_i;[\mb]_j}$, where $[n]_i$ denotes a particular partition of $n$, (i.e. $\{n_1,n_2,...\}:n_1+n_2 +...=n$),  are given in III. In Section 5 we will need it for the level four fields.

\section{Outline}

We will give an outline of the steps involved in this section so that the reader does not lose the woods for the trees. 

\begin{enumerate}
\item {\bf Step 1: In this approach, the action is made coordinate invariant and yet independent of $g_{\mu\nu}^R$}.  In III a world sheet action was written down that gave gauge invariant  equations of motion in the form of the ERG. We need to make this action invariant under general coordinate transformations. This was done in some detail for the graviton. We work out the technicalities for the massive fields in this section.This will be done by covariantizing derivatives using the reference metric $g_{\mu\nu}^R$. If the metric is flat it is easy to see that covariantising derivatives does not introduce any dependence on the background metric because of the relation $D_{\xn}D_{\xm}=D_{x_{n+m}}$.

\item {\bf Step 2: Regulating the world sheet theory:} We also include regulator terms and make sure they are invariant too in the same way. This involves adding higher derivative terms to the kinetic term. To make sure it is coordinate invariant we include the same metric $g_{\mu\nu}^R$. Schematically these terms are:
\[
\sum _{n} \int d^2z ~g_{\mu\nu}^R a_{n;\bar n}Y_n^\mu Y^\nu_\nb 
\]
These can be subtracted for the interaction Lagrangian where it modifies the massive field $S_{\mu\nu,n,\nb}Y_n^\mu Y_\nb ^\nu$.

\item 

{\bf Step 3: Modifying the tensor transformation laws :}  The transformation law of the massive fields are now modified by appropriate non tensorial terms to make the entire term coordinate invariant. Thus for instance we end up with combinations of the form
\[
(-S_{\mu \nu}  -a_{n,\nb} h_{\mu\nu}^R)Y_n^\mu Y_\nb^\nu\equiv \tilde S_{\mu\nu}Y_n^\mu Y_\nb^\nu
\]
Thus we will modify the transformation law for $S_{\mu\nu}$ with non tensorial terms coming  such that $\tilde S_{\mu\nu}$ is a tensor under coordinate transformations.

As a result of Steps 2 and 3 we have an action that is coordinate invariant, and   independent of $g^R_{\mu\nu}$. 
The physical quantities cannot depend on this background metric, and one expects that the solutions to the fixed point conditions will only depend on the quantity $g_{\mu\nu}^R + \tilde h_{\mu \nu} = g_{\mu\nu}$, since the original action only depended on $g_{\mu\nu}$. There are some subtleties here though. The functional measure ${\cal D}X(z)$ will have a metric dependence ${\cal D}X(z) \sqrt {g^R(X(z))}$ if we want the quantum theory to have manifest BGCT. This shows up in the  ERG, which introduces explicitly a background metric dependence through the two point function. This affects the intermediate equations but should not affect the solutions of the fixed point conditions or the on-shell S-matrix. We do not  have a proof of this. In quantum field theory there are proofs that the on-shell S-matrix is independent of the choice of background fields \cite{Abbott,DeWitt,Kallosh,GNW}. This must be true order by order in loops and thus must be true for the classical theory by itself. Since what we have is a classical (in space time) theory it should apply here also. We have not attempted to show this. 

\item {\bf Step 4: ERG and OPE: } We write down the exact RG. As shown in I,II and III this involves two terms. One is the free linear equation and the second is the quadratic interaction term. The quadratic term is in the form of a product of two gauge invariant field strengths at two points on the world sheet and one has to perform an OPE to write it at one point. This involves a Taylor expansion.
(It is also useful to point out that once we have gauge invariant equations we can set all the $\xn, \xnb $ to zero and work with ordinary vertex operators. This simplifies the Taylor expansion.)
The Taylor expansion of a scalar can be done in terms of covariant tensors (in {\em any} coordinate system). Let us refer to the RNC coordintes as $\bar Y^\mu$. The object we have in the ERG, is schematically of the form, in the RNC,
\[
\int dz_1 dz_2~\langle \bar Y^\mu (z_1)  \bar Y^\nu(z_2)\rangle {\delta {\cal L}\over \delta  \bar Y^\mu(z_1)}{\delta {\cal L}\over \delta \bar Y^\nu(z_2)}
\]
The product $ \bar Y^\mu (z_1){\delta {\cal L}\over \delta \bar Y^\mu(z_1)}$ is a scalar  if $\bar  Y^\mu$ is a geometric object. In the RNC,  it is a tangent vector at the origin, say O, of the RNC (which is $\bar Y^\mu=0$). In fact as we will see in Sec 7.3 it can equally well be taken to be a tangent vector at the point $\bar Y^\mu$, (call it P), since in the RNC the tangent vector is constant along a geodesic. Thus we will keep this interpretation for $\bar Y^\mu$ in the above equation, i.e. treat it as a (tangent to the geodesic from O) vector at the general point, P.  In any other (non RNC) coordinate system, say $Y^\mu$, we will still take it to be the {\em tangent vector to the geodesic at P} (call it $y^\mu$). In a general coordinate system it is an appropriately rotated version - $y^\mu= {\p Y^\mu\over \p \bar Y^\nu}|_{P}\bar Y^\nu$.  $ {\delta {\cal L}\over \delta  \bar Y^\mu(z_1)}$ is also a vector at P. Thus in a general coordinate system, the product $y^\mu(z_1) {\delta {\cal L}\over \delta  Y^\mu(z_1)}$ is a scalar. A detailed discussion appears in Sec 7.

\item {\bf Step 5: Dimensional Reduction:} In the loop variable approach the D+1st dimension, denoted by $\theta$, is assumed to be 
compact. The conjugate generalized momentum was denoted $q_n$. $q_0$ in particular was set equal to the mass of the field. $q_1$ is not physical and "$q$-rules" were defined (in I and II) to get  rid of $q_1$ without violating gauge invariance. For closed strings one requires that $q_1$'s occur in the combination $q_1\bar q_1$. The extra $q_1$'s can be removed by the same $q$-rules.

\end{enumerate}
\section{Step 1:General Covariance}
\subsection{Coordinate Transformations}

General Coordinate Invariance under coordinate transformations $X^{'\mu}= X^{'\mu}(X^\nu)\equiv X^\mu - \eps ^\mu(X)$ is what one demands in theories of gravity. Of course closed strings contain many more symmetries but general coordinate invariance is what
one would like manifest in the low energy theory.  \footnote{The consequences of the higher symmetries for the low energy theory is not understood.} Under this transformation $\p_zX^\mu$ and $\p_{\bar z}X^\mu$ transform as vectors. 
\[
\p_z X^{'\mu} = \p_z X^\nu {\p X^{'\mu}\over \p X^\nu}
\]
However $\pp X^\mu$ does not. Thus naively the string world sheet with massive backgrounds included is not general coordinate invariant under the {\em usual} tensor transformation laws for massive fields. If one modifies the transformation laws and combines it with gauge transformation laws it is presumably possible to make the action invariant. We will not follow this approach here. Instead we
will define covariant derivatives in the next section.

Another issue is that in the loop variable formalism we work with $Y^\mu$ rather than $X^\mu$, where $Y$ is a linear combination
of $X$ and its derivatives as given in \eqref{Y}. It may be a complicated problem to see what general coordinate transformations do to
$Y$, but we will cut the Gordian knot by pretending that our target space manifold is parametrized by $Y^\mu$ and demand general coordinate invariance under the coordinate change $Y^{'\mu}= Y^{\mu}(Y)$.  As before $Y_{n}^\mu= {\p Y^\mu \over \p \xn}$ is a vector on the tangent manifold. However $\pp Y^\mu \over \p \xn \p \xm$ is not and therefore ${\pp Y^{'\mu} \over \p \xn \p \xm} \neq {\p Y^{'\mu}\over x_{n+m}}$ in general, even if $Y^\mu$ satisfies this. Our strategy will be to construct a covariant version of ${\p \over \p \xn}$, ${D~~\over D_{\xn}}$. Gauge invariance would be satisfied if they obeyed $D_{\xn} D_{\xm} = D_{x_{n+m}} $. 

So, to summarize, we let $Y^{'\mu}= Y^{\mu}(Y)$ be the coordinate transformations under which ${\p Y^\mu\over \p \xn},~~n>0$ are vectors. 

\subsection{Covariant Derivatives}

An obvious candidate for the covariant derivative is:
\be	\label{CovDer}
{D\over D \xm} Y^\mu_n = {\p \over \p \xm}Y^\mu_n + \Gamma^\mu _{\rho \sigma} Y_m^\rho Y_n^\sigma
\ee
where $\Gamma ^\mu _{\rho \sigma}$ is the usual Christoffel connection. Under coordinate transformations it obeys:
\be  \label{Gamma}
\Gamma ^{'\nu '}_{\alpha ' \beta '} {\p Y^\la \over \p Y ^{'\nu'}}{\p Y^{'\alpha '}\over \p Y ^{\rho}}{\p Y^{'\beta '} \over \p Y ^{\sigma}}+
{\pp Y^{'\nu '}\over \p Y^\rho \p Y^\sigma} {\p Y^\la \over \p Y ^{'\nu'}}= \Gamma ^\la _{\rho \sigma}
\ee
Let us consider
\[
{D'\over D\xm}Y_n^{'\mu'} = {\p \over \p \xm} Y_n^{'\mu'} + \Gamma _{\alpha ' \beta'}^{'\mu'} Y_m^{'\alpha'}Y_n^{'\beta'}
\]
\[=
{\p \over \p \xm}({\p Y^{'\mu '} \over \p Y ^{\rho}} Y_n^\rho) + \Gamma _{\alpha ' \beta'}^{'\mu'}{\p Y^{'\alpha'} \over \p Y ^{\rho}} Y_m^\rho
{\p Y^{'\beta '} \over \p Y ^{\sigma}} Y_n^\sigma
\]
\[=
{\p Y^{'\mu '} \over \p Y ^{\rho}}{\p \over \p \xm}( Y_n^\rho) +{\pp Y^{'\mu '} \over \p Y ^{\rho}\p Y^\sigma}Y^\sigma_m Y^\rho_n
+ \Gamma _{\alpha ' \beta'}^{'\mu'}{\p Y^{'\alpha'} \over \p Y ^{\rho}} 
{\p Y^{'\beta '} \over \p Y ^{\sigma}} Y_m^\rho Y_n^\sigma
\]
\[=
{\p Y^{'\mu '} \over \p Y ^{\la}}[{\p \over \p \xm}( Y_n^\la) + {\p Y^\la \over \p Y ^{'\nu'}}[\Gamma _{\alpha ' \beta'}^{'\nu'}{\p Y^{'\alpha'} \over \p Y ^{\rho}} 
{\p Y^{'\beta '} \over \p Y ^{\sigma}} Y_m^\rho Y_n^\sigma +{\pp Y^{'\nu '} \over \p Y ^{\rho}\p Y^\sigma}Y^\sigma_m Y^\rho_n]
\]
\[=
{\p Y^{'\mu '} \over \p Y ^{\la}}[{\p \over \p \xm}( Y_n^\la)+ \Gamma ^\la_{\rho \sigma}Y^\sigma_m Y^\rho_n]={\p Y^{'\mu '} \over \p Y ^{\la}}
{D\over D \xm} Y_n^\la
\]
Thus we have a covariant version of $\p\over \p \xn$. (Note however that  $Y_n^\mu ={DY^\mu \over D\xn}\equiv {\p Y^\mu \over \p \xn}$ uses the ordinary derivative.) 

We should point out that the Christoffel connection $\Gamma$ used above should actually be called $\Gamma ^{R\mu}_{\nu \sigma}$ where $R$ stands for "reference" (or background). In III we had introduced a reference or background metric using which the kinetic and interaction terms were separately invariant under background coordinate transformations, i.e. the general covariance involved the
background metric, not the physical metric. Thus the physical metric occurred at most quadratically  whereas the background metric
occurred non polynomially. We will see this in more detail in the next section.

We have seen that one of the requirements for gauge invariance is that operators of the form $D_{x_n} D_{x_m}Y^\mu$ and also
$D_{x_n} D_{\bar x_m}Y^\mu$ need to be added to the action. Use of the covariant derivative ensures that these are all tensors under general coordinate transformations.

Similarly one can define its action on tensors:
\[
{D\over D\xn} \phi^\mu(Y)= Y_n^\rho \nabla _\rho^R \phi^\mu(Y)
\]
Here $\nabla^R$ is the background covariant derivative involving $\Gamma^R$.

One can check the consistency of these definitions by checking for instance its action on scalars formed out of vectors:  ${\p \over \p x_m}(\bar \phi_\mu(\bar Y)\bar Y_n^\mu)={\p \over \p x_m}(\phi_\mu(Y)Y_n^\mu)= {D \phi _\mu(Y)\over D\xm} Y_n^\mu + \phi_\mu (Y) {D Y_n^\mu \over D \xm}$. The bar denotes RNC.

\subsection{Implementing $D_{\xn} D_{\xm} Y^{\mu}= {D\over D x_{n+m}}Y^\mu=Y^{\mu}_{n+m}$}

Let us choose as canonical, the coordinate $\bar Y^\mu$ that was defined in Section 2 and obeys ${\pp \bar Y^{\mu} \over \p \xn \p \xm} = {\p \bar Y^{\mu}\over \p x_{n+m}}$. Let $Y^{'\mu}(\bar Y)$ be any other coordinate system. Then $Y_n ^{'\mu} = {\p Y^{'\mu}\over \p \bar Y^\rho} \bar Y_n^\rho$.
\be \label{3.12}
{D'\over D \xm} Y^{'\mu}_n = {\p Y^{'\mu}\over \p \bar Y^\rho} {D\over D \xm}\bar Y_n^\rho={\p Y^{'\mu}\over \p \bar Y^\rho}[\bar Y^\rho_{n+m}+ \bar \Gamma ^{R\rho}_{\alpha \beta} \bar Y^\alpha_m \bar Y^\beta _n]
\ee
On the other hand
\be  \label{3.13}
Y^{'\mu}_{n+m} = {\p Y^{'\mu}\over \p \bar Y^\rho} \bar Y^\rho_{n+m}
\ee
Clearly \eqref{3.12} and \eqref{3.13} can be equal only if $\bar \Gamma ^{R\mu}_{\rho\sigma}=0$. Thus in our canonical coordinate system the Christoffel connection should be zero, i.e. $g_{\mu\nu}^R = \eta_{\mu\nu}$.  This implies that in a general coordinate system where the background metric $g_{\mu\nu}^R \neq \eta_{\mu\nu}$, $\Gamma^R \neq 0$, but it should be flat, i.e. the curvature tensor should be zero. This was in fact the choice made in III. This is a limitation of this method. It is possible that there are other less restrictive ways of achieving gauge invariance.  It is important to note that this in no way restricts the manifold to be flat for the {\em physical} metric. But it does remove the possibility of choosing the background metric to be equal to the physical metric at the end of the calculation - which is often a useful trick in the background field formalism \cite{Abbott}.
Also the method is no longer background independent.

\subsection{Loop Variable using Covariant Derivatives}

\subsubsection{Massless mode vertex operator}

In III it was shown that the combined requirements of gauge invariance and masslessness of the graviton forced us to modify the Abelian gauge transformation and combine it with a non Abelian "tensor rotation" which in fact results in general coordinate transformations.

We review the construction given there (with some slight modification).

Our starting point is the action
\be	\label{massless}
S= \int d^2z~[\hf \eta_{\mu\nu} Y^\mu_1 Y^\nu_{\bar 1} - k_{1\mu} k_{\bar 1\nu} Y^\mu_1 Y^\nu_{\bar 1}\e + i K_{1;\bar 1\mu} Y_{1;\bar1}^\mu \e+ (massive)]
\ee
It is assumed that $\langle  -k_{1\mu} k_{\bar 1\nu} \e\rangle = \hf h_{\mu\nu}(Y)$ is the physical graviton fluctuation. (We set the antisymmetric part $B_{\mu\nu}=0$ to for simplicity. It is included in III.)  So $\eta_{\mu\nu}+h_{\mu\nu}=g_{\mu\nu}$ is the physical metric of space-time. $\langle i K_{1;\bar 1\mu}\rangle= S_\mu$ was identified in III as an auxiliary field necessary for gauge invariance of the interacting term. It was then shown that if the graviton is to be massless such a field
should not be there. It was then identified with a Christoffel connection for a background or reference metric $g^R_{\mu\nu}=\eta_{\mu\nu}+h^R_{\mu\nu}$ which
is introduced at intermediate stages. The final answer should not depend on $h^R_{\mu\nu}$. The gauge transformation was modified to include an action on the coordinates and $h^R_{\mu\nu}$ - background general coordinate transformations (BGCT). This was shown to be a symmetry of the EOM.

In more detail:

The {\bf gauge transformation} for $h_{\mu\nu}$ is
\be	\label{gthmunu}
\delta h_{\mu \nu} = \p_{(\mu} \xi_{\nu)}
\ee
The {\bf background general coordinate transformation} (BGCT) is defined to be the following:
\be\label{BGCThR}
\delta Y^\mu = -\eps ^\mu ;~~~~~\delta h_{\mu\nu}^R = \eta _{\rho \nu}\eps ^\rho_{~,\mu} + \eta _{\rho \mu}\eps ^\rho_{~,\nu} +
h_{\mu\nu,\rho}^R \eps^\rho + \eps ^\rho _{~,\mu}h_{\rho\nu}^R + \eps ^\rho_{~,\nu}h_{\mu\rho}^R
\ee
\be \label{BGCTh}
\delta h_{\mu\nu} = h_{\mu\nu,\rho} \eps^\rho + \eps ^\rho _{~,\mu}h_{\rho\nu} + \eps ^\rho_{~,\nu}h_{\mu\rho}
\ee
$h_{\mu\nu}$ transforms as an ordinary tensor under BGCT.    If we let $\eta _{\rho \nu}\eps^\rho = \xi_\nu$, then under the {\em combined}
action of BGCT and a gauge transformation, $h_{\mu\nu}$ transforms as it would under a GCT. Furthermore under this combined action
$h_{\mu\nu}-h^R_{\mu\nu}$ transforms as a tensor. It is this combined transformation, which we will refer to as "massless gauge invariance" from now on,  that will be a manifest symmetry of the theory and also thus of each EOM. Since we will ensure that the full action - and therefore the continuum physics - does not depend on $h_{\mu\nu}^R$, this "gauge invariance" is equivalent to GCT invariance of the theory. In addition to this the massive fields have their usual "massive gauge invariances".

All massive field tensors are initially assumed to have their usual tensorial transformations under BGCT. This will be modified later in Section 6 (Step 3).
 
Thus the massless action is modified to
\be \label{massless1}
S= \int d^2z~[\hf (\eta_{\mu\nu} +h_{\mu\nu}^R)Y^\mu_1 Y^\nu_{\bar 1} - [k_{1\mu} k_{\bar 1\nu}  -\hf  k_{0(\nu}K_{1;\bar 1\mu)}] Y_{1}^\mu Y^\nu_{\bar 1}\e
\ee
We have integrated by parts on $\bar x_1$. 
The combination in square brackets $(k_{1\mu} k_{\bar 1\nu}  -\hf  k_{0(\nu}K_{1;\bar 1\mu)})Y_{1}^\mu Y^\nu_{\bar 1}\e$ was gauge invariant under the loop variable transformation. We set it equal to (or replace it with)
$(h_{\mu\nu}-h_{\mu\nu}^R)Y_{1}^\mu Y^\nu_{\bar 1}\e$.

Now  under this combined action of gauge transformation plus BGCT $(h_{\mu\nu}-h_{\mu\nu}^R)$ transforms as a tensor and the action \eqref{massless1} is invariant under this (combined action of gauge transformations and BGCT).

Thus \eqref{massless1} becomes:
\be \label{massless2}
S= \underbrace{\int d^2z~[\hf (\eta_{\mu\nu} +h_{\mu\nu}^R(Y))Y^\mu_1 Y^\nu_{\bar 1}}_{K} + \underbrace{\int d^2z~\hf (h_{\mu\nu}(Y)-h^R_{\mu\nu}(Y)) Y_{1}^\mu Y^\nu_{\bar 1}}_{S_{int}}]
\ee

As promised the full action has no dependence on $h^R_{\mu\nu}$. But $S_{int}$ and $K$ separately do have this dependence and so the EOM does.  As was shown in III (and is repeated below) the result is that the  interaction term in the graviton equation, which is a product of gauge invariant "field strengths", has gauge invariance when the gauge transformation is accompanied by background GCT.

\subsubsection{{Normal Coordinates}}

Let O be the origin of our RNC, with coordinates $x_0$ and P be a general point with coordinates $x$. As in \cite{AGFM,Pet,Eis} we consider geodesics starting from O and going through P and denote by $\vec \xi$ the unit tangent to the geodesic at O. This is a geometric object. We let $t$ be the proper distance from O to P along this geodesic. It can be shown that \cite{Eis}
\be    \label{NC}
x^\mu = x_0^\mu + t \xi ^\mu - {t^2\over 2!} \xi ^\rho \xi ^\sigma \Gamma ^\mu _{~ \rho \sigma} - {t^3\over 3!} \xi ^\rho \xi ^\sigma \xi ^\lambda \Gamma ^\mu _{~\rho \sigma \la} +...
\ee
where the $\Gamma$ with $n$ indices is recursively defined in terms of derivatives and products of $\Gamma$ with $n-1$ indices. 

Normal coordinates $\bar Y^\mu$ are introduced by defining $\bar Y^\mu =t \xi^\mu$. Thus on the one hand $\bar Y^\mu$ is a geometric object - a vector at O. On the other hand it is also a coordinate, related to $x$ by
\be
x^\mu = x_0^\mu +\bar Y ^\mu - {1\over 2!} \bar Y ^\rho \bar Y^\sigma \Gamma ^\mu _{~ \rho \sigma} - {1\over 3!} \bar Y ^\rho \bar Y ^\sigma \bar Y^ \lambda \Gamma ^\mu _{~\rho \sigma \la} +...
\ee

Our kinetic term K, which is a coordinate scalar at $x$ is unchanged when  written in RNC:
\be	\label{KT}
g_{\mu\nu}(x) \p_z x^\mu \p_\zb x^\nu = \bar g_{\mu \nu}(\bar Y) \p_z \bar Y^\mu \p_\zb \bar Y^\nu 
\ee

In RNC, the metric tensor  has a Taylor series expansion:
\be	\label{metric}
\bar g_{\mu\nu}(\bar Y) = \bar g_{\mu \nu}(0) - {1\over 3} \bar Y^\al \bar Y^\beta  \bar (R_{\mu \al \nu \beta}(0) )....
\ee

Note that the LHS is a tensor at P.  The RHS is a sum of tensors at O provided under coordinate transformations $\bar Y^\mu$ is transformed as a vector at O (and not just as a coordinate).  Note that the two sides are thus tensors at {\em different} points and  transform differently under coordinate transformations. Thus this equation, as it stands, is only valid in the RNC.

Thus the kinetic term \eqref{KT} becomes:
\be  \label{KT1}
\bar g_{\mu \nu}(\bar Y) \p_z \bar Y^\mu \p_\zb \bar Y^\nu= \bar g_{\mu \nu}(0)\p_z \bar Y^\mu \p_\zb \bar Y^\nu - {1\over 3} \bar Y^\al(z) \bar Y^\beta (z) \bar R_{\mu \al \nu \beta}(0)\p_z \bar Y^\mu \p_\zb \bar Y^\nu +....
\ee
In the LHS one has to interpret $\bar Y^\mu$ as a coordinate and $\p_z \bar Y^\mu$ as a vector at P. In the RHS on the other hand, one has to interpret $\bar Y^\mu$ as a vector at O.  A covariant derivative of a vector $A^\mu$ at the point $x$ in general would be defined as
\[
D_z A^\mu = \p_z A^\mu +  \Gamma ^\mu_{~\rho \sigma} (x)\p_z x^\rho  \p_z A^\sigma
\]
Since $\bar \Gamma ^\mu _{~\rho \sigma}|_{\bar Y=0} =0$ (being in a RNC), $\p_z = D_z$ and thus $\p_z \bar Y^\mu$ can also be treated as a vector at O. In this case both sides of \eqref{KT1} are scalars (LHS at P and RHS at O) and the equation is valid in any coordinate system, provided we transform the
terms appropriately.  Thus, transforming from RNC back to the $x$ coordinate \eqref{NC}, let us define the geometric object,
$y^\mu|_O$ by 
\[ 
y^\mu|_O = {\p x^\mu \over \p \bar Y^\nu}|_{\bar Y=0}\bar Y^\nu= \delta ^\mu_{~\nu} \bar Y^\nu = \bar Y^\mu = t\xi^\mu
\] Then we get
\be  \label{KT2}
 g_{\mu \nu}(x) \p_z  x^\mu \p_\zb  x^\nu=  g_{\mu \nu}(x_0)D_z y^\mu|_O D_\zb y^\nu|_O - {1\over 3}y^\al(z)|_O y^\beta (z)|_O  R_{\mu \al \nu \beta}(x_0)D_z y^\mu|_O D_\zb y^\nu|_O +....
\ee

Here $D_zy^\mu$ is the covariant derivative at O: $D_zy^\mu|_O = \p_z y^\mu|_O + \Gamma ^\mu_{~\rho \sigma}(x_0) \p_z x^\mu|_O y^\sigma|_O$.
We will work throughout in the RNC using \eqref{KT1} for simplicity and covariantize at the end.

In \eqref{KT1} the second term quartic in $\bar Y^\mu$ should be treated as part of the interaction and will be included in $S_{int}$. Thus 
\[
S_{int} = \int d^2z ~[ - {1\over 3} \bar Y^\al(z) \bar Y^\beta (z) \bar R_{\mu \al \nu \beta}(0)+ O(\bar Y^3)]Y_{1}^\mu Y^\nu_{\bar 1}
\]
However in this paper we consider only flat backgrounds and terms involving the curvature tensor will be dropped.

\subsubsection{{ Field Strength}}
Let us turn to the equations of motion as given by the ERG \eqref{ERG} and the expression for the functional derivative \eqref{FD}:
The functional derivative below gives the field strength:
\[	
{\delta S_{int} \over \delta \bar Y^\rho(z)} - {\p \over \p x_1}{\delta S_{int} \over \delta \bar Y_1^\rho(z)}- {\p \over \p \bar x_1}{\delta S_{int} \over \delta \bar Y_{\bar 1}^\rho(z)}=
\]
\[
  \hf[{\p \bar {\tilde h}_{\mu \nu}\over \p \bar Y^\rho}(\bar Y(z))  - {\p \bar {\tilde h}_{\rho \nu}\over \p \bar Y^\mu}(\bar Y(z))-{\p \bar {\tilde h}_{\mu \rho}\over \p \bar Y^\nu}(\bar Y(z))
+{2\over 3} \bar Y^\beta(z) (\bar R^R_{\rho \nu \mu \beta}(0)+\bar R^R_{\rho \mu \nu \beta}(0))]\bar Y^\mu_1(z) \bar Y^\nu _{\bar 1}(z)\]
\be	\label{GravFS}
+[ \bar {\tilde h}_{\rho \mu}(\bar Y(z))+{1\over 6} \bar Y^\alpha(z) \bar Y^\beta (z)(\bar R^R_{\rho \alpha \mu \beta}(0)+\bar R^R_{\mu \alpha \rho \beta}(0))]\bar Y^\mu_{1;\bar 1}(z)
\ee
where $\bar{\tilde h}_{\mu\nu}= \bar{h}_{\mu \nu}-\bar {h}^R_{\mu\nu}$ and all arguments of fields  have been displayed to avoid confusion. The bars on the metric fluctuation and curvature tensor are just to remind us that we are working in the RNC. To go to a general coordinate system we just remove the bars.
The field strength tensor is, to this order,
\be \label{CovGravFS2}
F_{\rho \mu \nu}(\bar Y)= \hf[{\p \bar {\tilde h}_{\mu \nu}\over \p \bar Y^\rho}(\bar Y(z))  - {\p \bar {\tilde h}_{\rho \nu}\over \p \bar Y^\mu}(\bar Y(z))-{\p \bar {\tilde h}_{\mu \rho}\over \p \bar Y^\nu}(\bar Y(z))
+{2\over 3} \bar Y^\beta(z) (\bar R^R_{\rho \nu \mu \beta}(0)+\bar R^R_{\rho \mu \nu \beta}(0))+...]
\ee

The field $\tilde h$, and the curvature tensor are gauge covariant and thus so is the field strength. 

 We can write it in a general coordinate system by   the usual procedure  of writing background covariant derivatives: 
\be  \label{CovGravFS}
(\Gamma_{\rho \mu \nu }-\Gamma^R_{\rho \mu \nu })\rightarrow  \hf(\nabla ^R _\mu \tilde h_{\rho \nu} + \nabla^R _\nu \tilde h_{\rho \mu}  - \nabla ^R _\rho \tilde h_{\mu\nu})\equiv\tilde \Gamma^R_{\rho \mu \nu} 
\ee

The curvature tensors are manifestly covariant already - the bars just need to be removed.

The quadratic term in the EOM is an (Operator) product of field strengths at different locations on the world sheet. Thus the above expression has to be Taylor expanded in powers of $z$. This is discussed in Section 7.

In this paper we set the curvature tensor to zero, so only the terms involving $\tilde h_{\mu\nu}$ in \eqref{CovGravFS2} need be kept.

\subsubsection{{ Free Equation}}

One can also calculate the free equation which involves second order derivatives:

The constraint $K_{1;\bar 1}.k_0=k_1.k_{\bar 1}$ (this "K-constraint" requirement was derived in an Appendix of III) becomes 
\be	\label{Constr}
h^\mu_{~\mu} - h^{R\mu}_{~~\mu}=\tilde h^\mu_{~\mu}=0
\ee
(Note that the index $\mu$ runs over  $D+1$ values from 0 to $D$ and thus $\kim k^\mu_1$ is actually $ \kim k^\mu_1 + q_1 q_1\to h^\mu_{~\mu} + \Phi_D$ and includes the dilaton.)
The free graviton equation for the metric fluctuation was derived in III and is:
\be 	\label{GravFree}
\p^\rho (\Gamma_{\rho\mu\nu}-\Gamma^R_{\rho\mu\nu})=0
\ee
 \eqref{GravFree} is the RNC version of 
the covariant equations in a general coordinate system, at the origin, where $\Gamma^R=0$:
\be   \label{GravFreeCov1}
\nabla^{R}_{\sigma}( g^{R\sigma \rho}\tilde \Gamma_{\rho\mu\nu})=0
\ee

We  work out for completeness the contribution due to the rest of the terms involving the background curvature tensor. In this paper we set this contribution to zero. The equation in loop variable notation is
 \be
\hf[ -\ko^2 \kim \kinb + \ko .\ki (\kom \kinb + \kon \kimb) - \kom \kon \ki .\kib] =0
 \ee
 The equation is being evaluated at the origin O, where $\bar Y^\mu=0$, so only the quadratic term contributes - the cubic and higher order terms do not contribute.

Therefore  we use 
 \[ 
 \kim \kinb = {1\over 6} \bar Y^\al \bar Y^\beta    ( \bar {R}^R_{\mu \al \nu \beta}(0) +\bar R^R_{\nu \al \mu \beta}(0))
 \]
 in the above and obtain
 \be
 - \bar R^R_{\mu \nu} \bar Y_1^\mu \bar Y_{\bar 1}^\nu
 \ee
 Thus the total for the graviton contribution  to the free graviton EOM is (dropping bars):
 \be \label{GravFreeCov}
(  R^R_{\mu \nu}+\nabla^{R}_{\sigma}( g^{R\sigma \rho}\tilde \Gamma_{\rho\mu\nu}))  Y_1^\mu  Y_{\bar 1}^\nu
\ee

\subsubsection{Comparison with Einstein's Equation}

This free equation in the first case should be compared with what one expects for a graviton from Einstein's vacuum equation $R_{\mu\nu}=0$ expanded to linear order in $\tilde h$, 
about a background. One can expand as follows:
\be   \label{Eins}
R_{\mu \nu} = R^R_{\mu \nu} + \delta R_{\mu \nu}
\ee

To evaluate $\delta R$,
 go to an inertial frame with $\Gamma =0$ at the point under consideration,
\[R^\alpha _{~\mu \beta \nu} = \p_\beta \Gamma ^\alpha _{~\mu \nu} - \p_\nu \Gamma ^\alpha _{~\mu \beta}
\]
So
\[\delta R^\alpha _{~\mu \beta \nu} = \p_\beta \delta \Gamma ^\alpha _{~\mu \nu} - \p_\nu \delta \Gamma ^\alpha _{~\mu \beta}
\]
Now unlike $\Gamma$, $\delta \Gamma ^\alpha _{~\mu \nu}$ is a tensor, so the above equation, if written covariantly, is valid in all frames:
\be 
\delta R^\alpha _{~\mu \beta \nu} = \nabla_\beta \delta \Gamma ^\alpha _{~\mu \nu} - \nabla_\nu \delta \Gamma ^\alpha _{~\mu \beta}
\ee
So we get the Palatini equation:
\be 	\label{Palatini}
\delta R_{\mu \nu}=\delta R^\alpha _{~\mu \alpha \nu} = \nabla_\alpha \delta \Gamma ^\alpha _{~\mu \nu} - \nabla_\nu \delta \Gamma ^\alpha _{~\mu \alpha}
\ee
We now show that to linear order in $\tilde h$ (or $h$), 
\be   \label{Result}
\delta \Gamma ^\alpha _{~\mu \nu}\equiv\Gamma ^\alpha _{~\mu \nu}-\Gamma ^{R\alpha} _{~\mu \nu}=g^{R \alpha\rho}\tilde \Gamma _{\rho \mu \nu}
\ee
Writing $g^{\rho \sigma}= g^{R\rho \sigma}+\delta g^{\rho \sigma}$ we get 
\[
\Gamma ^\sigma _{~\mu \nu} -\Gamma ^{R\sigma} _{~\mu \nu}= g^{R\rho \sigma}(\Gamma _{\rho \mu \nu} -\Gamma^R _{\rho \mu \nu})+\delta g^{\rho \sigma} \Gamma _{\rho \mu \nu}
 \]
\be \label{LHS}
=g^{R\rho \sigma}(\Gamma _{\rho \mu \nu} -\Gamma^R _{\rho \mu \nu})+\delta g^{\rho \sigma} \Gamma^R _{\rho \mu \nu}
\ee
to linear order in $h$.

Now consider the RHS of \eqref{Result}. Expand the covariant derivatives:
\be \label{RHS}
g^{R\rho \sigma} [(\Gamma _{\rho \mu \nu} - \Gamma^R _{\rho \mu \nu})-\Gamma^{R\alpha}_{~\mu \nu} \tilde h _{\rho \alpha}
\ee

If we now take into account the fact that $\delta g^{\rho \sigma}= -g^{R\rho \alpha}h_{\alpha \beta}  g^{R\beta \sigma}$
we see that \eqref{LHS} and \eqref{RHS} are equal and we have the result \eqref{Result}. Furthermore taking the trace we get
\be \label{Trace}
\Gamma ^\alpha _{~\mu \alpha}-\Gamma ^{R\alpha}_{~~\mu \alpha} = g^{R\alpha \rho}\hf(\nabla _\mu^R\tilde h_{\rho \alpha}+\nabla^R_\alpha \tilde h_{\mu \rho} -\nabla ^R_\rho\tilde h_{\mu \alpha})=0
\ee
We have used the constraint that $h^{R\mu}_{~\mu}=h^{\mu}_{~\mu}$ or $\tilde h^\mu_{~\mu}=0$. We set $\Phi_D=0$ for convenience and let $\mu$ run from $0$ to $D-1$.

Inserting \eqref{Result} and \eqref{Trace} into \eqref{Palatini} we obtain the  equation for $\tilde h$ in the background metric (including for completeness the background contribution):
\be
R^R_{\mu \nu}+ \nabla^R_\sigma (g^{R\sigma \rho} \tilde \Gamma _{\rho \mu \nu} =0
 \ee
 which is the covariantized equation that we obtained, \eqref{GravFreeCov}.

\eqref{GravFreeCov1}, in the linearized (i.e. assuming that both $h$ and $h^R$are infinitesimal) approximation,  becomes
\be 	\label{Gravstandard}
\p^\rho \Gamma_{\rho\mu\nu} - \hf\p_\mu\p_\nu h^\rho_{~\rho} = 
\hf[-\p^2 h_{\mu\nu} + \p_\mu \p^\rho h_{\rho\nu} + \p_\nu \p^\rho h_{\mu \rho} -\p_\mu\p_\nu h^\rho_{~\rho}]=0
\ee
This is a standard form of the linearized graviton equation in flat space.
In writing \eqref{Gravstandard}, use has been made of the fact that in the linearized approximation $h_{\mu\nu}^R$ is of the form $ -\hf \p_{(\mu} \xi _{\nu)}$ and also that then
the constraint \eqref{Constr} becomes $\p^\rho \xi_\rho = h^\rho_{~\rho}$.

\subsubsection{Massive mode vertex operators for open strings and closed strings}

As explained in Section 2 we need vertex operators of the form $K_{\mu,n,m}{\pp Y^\mu\over \p \xn \p \xm}$ and also higher derivatives for open strings and for closed strings.   We simply replace them with our covariant derivatives. Thus for instance, we have 
\[
{\pp Y^\mu\over \p \xn \p \xm}\to {D^2 Y^\mu\over D \xn D \xm} 
\]
Since in our case, ${D^2 Y^\mu\over D \xn D \xm}={DY^\mu\over Dx_{n+m}}$ this expression does not in fact depend on the background metric. So covariantising does not introduce background dependence in the action.

\subsubsection{Mixed derivative vertex operators}
For closed strings, in the presence of a finite cutoff 
we need in addition mixed derivative terms ${\pp Y^\mu\over \p \xn \p {\bar x_m}}$. These are covariantized to  ${D^2 Y^\mu\over D \xn D {\bar x_m}}$.

The vertex operators ${D^2 Y^\mu\over D \xn D {\bar x_m}}$ do depend on $\Gamma ^R$ so we get in the loop variable the following terms:
\[
K_{n;\bar m \mu}{D^2 Y^\mu\over D \xn D {\bar x_m}}=K_{n;\bar m \mu}( {\pp Y^\mu\over \p \xn \p {\bar x_m}}+ \Gamma ^{R\mu}_{\rho \sigma}Y^\rho_n Y^\sigma_{\bar m})
\]
In our approach,
to cancel this,  we  subtract the term $\Gamma ^{R\mu}_{\rho \sigma}Y^\rho_n Y^\sigma_{\bar m}$  as follows:
 \be   \label{Nontensor}
 (-k_{n \rho} k_{\bar m \sigma }-iK_{n;\bar m \mu}\Gamma ^{R\mu}_{\rho \sigma})Y^\rho_n Y^\sigma_{\bar m}
 \ee
 If $n= m$ this vertex operator is a physical closed string mode and will be there in the Lagrangian. Thus 
\be \label{Nontensor1}
\langle   (-k_{n \rho} k_{\bar n \sigma }-iK_{n;\bar n \mu}\Gamma ^{R\mu}_{\rho \sigma}) = -S_{n,\bar n,\rho \sigma} - S_{\mu n,\bar n} 
\Gamma ^{R\mu}_{\rho \sigma}\equiv \tilde S_{n,\bar n,\rho \sigma}
\ee
and $\tilde S$ will be defined to have tensorial transformation property. 

The fact that ${D^2 Y^\mu\over D \xn D \xm}={DY^\mu\over Dx_{n+m}}$ is important.  It ensures that all the $K_{\mu[n_i];[\bar m_j]}$ with $\sum _i n_i=n$ and $\sum _j m_j=m$ introduce the same dependence  $\Gamma ^{R\mu}_{\rho \sigma}Y^\rho_n Y^\sigma_{\bar m}$. Then using (derived in III)
\[
\sum_{i,j}K_{\mu[n_i];[\bar m_j]}=K_{\mu n;\bar m}
\]
we see that the field redefinition in the free equation and in the interacting equation are the same.

\section{Step 2: Higher derivative kinetic terms} 

   What we have done so far is to write down a world sheet action that has GCT, but the kinetic term and interaction term are not separately invariant under GCT, but are separately invariant under Background GCT that includes transformation of the background metric $h^R_{\mu \nu}$. But in the ERG formalism we also have to worry about the regulated theory. We have to ensure that the regulator
   is consistent with these symmetries. A simple way to ensure this is the following: Add higher derivative terms of the form $\sum _{n,\bar m =1}^N{a_{n,\bar m} a^{n+\bar m-2}}(\eta_{\mu\nu}+h_{\mu\nu}^R)Y_n^\mu Y_{\bar m}^\nu $. Here $a_{n,\bar m}$ are some coefficients that will determine the precise nature of the cutoff. Explicit powers of the cutoff have also been introduced. The sum can be extended to infinity. This ensures convergent high energy behaviour. It is clear that
   this term is invariant under BGCT. But we have thus introduced a dependence on $h_{\mu\nu}^R$ albeit in the intermediate stages - once $a\to 0$, these terms disappear. Nevertheless we would like BGCT even in the intermediate stages. So in order to cancel
   the $h_{\mu\nu}^R$ dependence, we add the same terms with the opposite sign in the interaction Lagrangian. These are
   modifications of the massive vertex operators. Thus the coefficient of $Y_n^\mu Y_{\bar m}^\nu$ changes further and  \eqref{Nontensor} becomes 
   \be   \label{NT2}
   (k_{n\mu}k_{\bar m \nu}- K_{n;\bar m \alpha}\Gamma^{R\alpha}_{\mu \nu} - a_{n,\bar m}h_{\mu\nu}^R) Y_n^\mu Y_{\bar m}^\nu
   \ee
   In the case $n= m$,  the definition of $\tilde S$ becomes modified from \eqref{Nontensor1}:
\be  \label{Nontensor3}
\langle   (-k_{n \rho} k_{\bar n \sigma }-iK_{n;\bar n \mu}\Gamma ^{R\mu}_{\rho \sigma}- a_{n,\bar n}h_{\mu\nu}^R)\rangle = -S_{n,\bar n,\rho \sigma} - S_{\mu n,\bar n} 
\Gamma ^{R\mu}_{\rho \sigma} -a_{n,\bar n} h_{\mu\nu}^R\equiv \tilde S_{n,\bar n,\rho \sigma}
\ee

\section{Step 3: Modifying Transformation Laws}

In the previous section we have seen that the massive mode vertex operators are modified by the addition of non tensorial objects involving the Christoffel connection $\Gamma^R$ and a piece of the background metric $h_{\mu\nu}^R$. This makes the action non invariant under BGCT. This is rectified by modifying the transformation laws of the massive fields. Thus
 $\Gamma ^{R\mu}_{\rho \sigma}$ is not a tensor. It's non tensorial transformation is 
 \[
 \delta \Gamma ^{R\mu}_{\rho \sigma}=-{\pp Y^{'\mu}\over \p Y^\rho \p Y^\sigma}=
- k_{0\rho} k_{0\sigma} \eps ^\mu
\] Here $\delta Y^\mu = Y^{'\mu}-Y^\mu = -\eps ^\mu$. 
Similarly $h_{\mu\nu}^R$ is also not a tensor.($\eps _\mu = \eta_{\mu\nu}\eps^\nu$)
\[
\delta h_{\mu\nu}^R = \underbrace{{\p \epsilon _\mu \over \p Y^\nu}+{\p \epsilon _\nu \over \p Y^\mu}}_{non~tensor} +\underbrace{ (\eps ^{R\la} h_{\mu\nu,\la} + \eps ^\la_{~,\mu}h^R_{\la\nu}+\eps^\la_{~,\nu}h^R_{\mu\la})}_{tensor~rotation}
\]
Thus the expression \eqref{Nontensor},\eqref{Nontensor1} and \eqref{NT2} are not invariant under background GCT. But we modify the transformation property of the physical field to cancel this offending non tensorial piece.

Thus 
\[
\delta_{non-tensorial} S_{n,\bar n,\rho \sigma}=-(S_{\mu n,\bar n}{\pp \epsilon^{\mu}\over \p Y^\rho \p Y^\sigma}+ a_{n,\bar n}({\p \epsilon _\rho \over \p Y^\sigma}+{\p \epsilon _\sigma \over \p Y^\rho}))
\]
 This makes the field $\tilde S$ in \eqref{Nontensor3} a tensor.

 We have thus made all the massive mode vertex operators (including mixed derivatives operators) covariant by adding and subtracting the Levi-Civita connection $\Gamma^{R\mu}_{\rho\sigma}$ in appropriate places in such a way that no dependence on $\Gamma^R$ is actually introduced in the final answer.

To summarize this section: 
We have replaced derivatives with background covariant derivatives and canceled the extra added pieces with corresponding terms of the opposite sign elsewhere in the action. We now have an action that is  independent of $g_{\mu\nu}^R$. Yet after modifying the transformation laws of the massive fields under a BGCT  to include some non tensorial terms, the theory is manifestly invariant under BGCT!  Invariance under BGCT in a theory that does not have $g_{\mu\nu}^R$ means that it is invariant under GCT.   Thus the solution to the equations of motion is also expected to be invariant under GCT.  The regulator terms involve arbitrarily high derivatives. Absorbing
the non tensorial terms in this into massive fields is possible because we have an infinite tower of massive fields in string theory.  

There is one important caveat: The action is independent of $g_{\mu\nu}^R$, yet the coefficients of covariant vertex operators separately have dependence on $g_{\mu\nu}^R$. The structure of the ERG (and the functional integration measure) also require a background metric.  Thus the individual equations of motions obtained from the ERG, that equate coefficients of covariant vertex operators will have dependence on $g_{\mu\nu}^R$. This is exactly the situation described in Appendix C of III where an example in Yang-Mills theory was described. Individually each equation depends on the background but one expects that the continuum physics does not. Thus if one solves for all the massive modes and obtains an equation for the graviton, this should not have any dependence on $g_{\mu\nu}^R$. One thus expects that the interactions will modify $h^R_{\mu\nu}$ to $h^R_{\mu\nu}+ \tilde h_{\mu\nu}= h_{\mu\nu}$, in the definition of $\tilde S$.
Similarly the solutions to these equations describe fixed points of the theory and should presumably be independent of $g_{\mu\nu}^R$. \footnote{As mentioned in Section 3, in quantum field theory there are proofs that the on-shell S-matrix is independent of the choice of background fields \cite{Abbott,DeWitt,Kallosh}.  Presumably these proofs apply here also, though
in string theory we  have an infinite number of fields, unlike in the theories studied by these authors.}
We have not attempted a proof of these statements.

\section{Step 4: ERG and Covariant OPE}

The ERG is reproduced here for convenience:
\subsection{ERG}
\[
\int du ~ {\p L[X(u)]\over \p \tau} =
\]
\be	\label{ERG1}
 \int dz~\int dz'~\hf~\dot G^{\mu \nu}(z,z')\Bigg( \int du~{\delta^2L[X(u)] \over \delta X^\nu(z')\delta X^\mu(z)}+
\int du~\int dv~ {\delta L[X(u)]\over \delta X^\mu(z)}{\delta L[X(v)]\over \delta X^\nu(z')}\Bigg)
\ee

Here $L[X(u)]$ is the two dimensional world sheet theory written in terms of loop variables. Thus 
\[
L[X(u)]=e^{i{\cal L}[X(u)]}
\]

The variables $z,X(z)$ are to be understood as generalized in \eqref{zxn}-\eqref{zxnb}. As shown in I,II and III, these equations are
gauge invariant. In fact each of the two terms in the ERG is gauge invariant. The first term gives the free gauge invariant equations of motion. The second term is thus a product of two gauge invariant "field strengths" at $z$ and $z'$. We have also seen in III (and summarized in Sec 4)  that in the closed string case, there is a subtlety that involves the massless mode. Making the "field strength" gauge invariant required us to include
coordinate transformations as part of the gauge transformation. This was achieved by introducing a fictitious reference metric. The BGCT
(general coordinate transformations involving the background metric) hold separately for the kinetic term in the action as well as the interaction term.

\subsection{Functional Derivatives}

  We have achieved general coordinate invariance
 by covariantizing derivatives. When evaluating functional derivatives as in \eqref{FD} the presence of $\Gamma$ in the covariant derivatives introduces some complication. We get around this by working in RNC. Since the action is completely invariant we can do this without loss of generality.   The final equations can be covariantized by reintroducing covariant derivatives and the fields $\tilde S$ in place of $S$.

\subsection{Kinetic Term and Green function}
 The kinetic term $S_0$ is covariantized by the introduction of $g_{\mu\nu}^R$. The ERG involves the two point function $G^{\mu\nu}(z,z') = \langle Y^\mu(z)Y^\nu (z')\rangle$ which computed using $S_0$. We can expect this term to be background covariant provided $Y^\mu$ is a tensorial object - which it is in the RNC where it is a geometric object - a vector tangent to the geodesic. Let us refer to it in the RNC as $\bar Y^\mu$. Thus we will assume that $G^{\mu \nu}(z,z')=\langle \bar Y^\mu(z) \bar Y^\nu (z')\rangle$ is defined in the RNC and in any other coordinate system it is obtained by transforming $\langle \bar Y^\mu(z) \bar Y^\nu (z')\rangle$ appropriately, i.e. as a tensor product of two vectors. (This is clearly not the same as $\langle Y^\mu(z)Y^\nu (z')\rangle$, which is not a tensor at all.)
 
 We elaborate on this idea: Let $X$ be a general coordinate system. At a point O (with coordinates $X_0$) we set the origin of an RNC system $\bar Y^\mu$. The point O has coordinate $\bar Y^\mu=0$. For a general point P with coordinate $X$, we consider a geodesic
 that starts from O and goes through P. Let the tangent vector to this geodesic at O be $\vec \xi _P$ and the proper distance along this geodesic to P be $t_P$. Then $\bar Y^\mu=t_P\xi_P ^\mu$. $\vec \xi_P$ is a geometric object - a vector  at O, {\em not} at P. So $\bar Y^\mu$ transforms as a vector at O. One would like an object that is a vector at P. So let us define the tangent vector field, $\xi^\mu(P)$ (or $\xi ^\mu (X_P)$) of unit norm vectors tangent to the geodesics through O at the (general) point P.  They obey
 \[
 \xi ^\nu \nabla _\nu \xi ^\mu = \xi ^\nu {\p \xi ^\mu \over \p X^\nu} + \Gamma ^\nu_{\mu \rho} \xi ^\mu \xi^\rho=0
 \]
 In the RNC this equation becomes
 \[
\bar \xi ^\nu \nabla _\nu \bar \xi ^\mu = \bar \xi ^\nu {\p \bar \xi ^\mu \over \p X^\nu} + \bar \Gamma ^\nu_{\mu \rho} \bar \xi ^\mu \bar \xi^\rho=0
\] 
But we know that in the RNC all along the geodesic,
\[
\bar \Gamma ^\nu_{\mu \rho} \bar \xi ^\mu \bar \xi^\rho=0
\]

This follows from the fact that $\bar Y^\mu$ satisfies the geodesic equation 
\[
{d^2\bar Y^\mu\over dt^2}+ \bar \Gamma ^\nu_{\mu \rho} {d \bar Y ^\mu\over dt} {d\bar Y^\rho\over dt}=0
\] 
and since ${d^2\bar Y^\mu\over dt^2}=0$ we get $ \bar \Gamma ^\nu_{\mu \rho} {d \bar Y ^\mu\over dt} {d\bar Y^\rho\over dt}=
 \bar \Gamma ^\nu_{\mu \rho} \bar \xi^\mu \bar \xi^\rho=0$.
  
Thus we get   
\[
\bar \xi ^\nu {\p \bar \xi ^\mu \over \p X^\nu}=0
\]
which means $\bar \xi^\mu$ is constant along a geodesic. A solution to this is thus the constant (along a geodesic) vector field $\bar \xi^\mu (\bar Y_P) = \bar \xi_P$. This is just the obvious fact that in the RNC geodesics are straight lines through the origin, so the tangent vector field is a constant (along a geodesic) vector field. We thus see that in the RNC $\bar Y^\mu_P = t_P\bar \xi^\mu (\bar Y_P) $ is not only a coordinate, it is also a vector field, i.e. the two objects coincide.  This will not be the case in a general coordinate system.

Thus when we change coordinates to $Y$, the vector field $\bar Y^\mu(\bar Y_P) = t_P \bar \xi^\mu(\bar Y_P)$ transforms like a vector field {\em at P} to a new vector field, 
\be 	\label{Vecfld}
y^\mu(P) = t_P \xi ^\mu (Y_P) = t_P{\p Y^\mu \over \p \bar Y^\nu}|_P \bar \xi^\nu (P)
\ee
 whereas the coordinate $\bar Y^\mu _P$ becomes $Y^\mu _P$.
 
  Now the Green function is $\langle \bar Y^\mu (z) \bar Y^\nu (z')\rangle $ in the RNC. We  {\em define} the corresponding Green function in a general coordinate system to be 
  \be \label{CovGF}
G^{\mu \nu}(z,z')\equiv  \langle y^\mu(z)|_P y^\nu (z')|_{P'}\rangle = {\p Y^\mu \over \p \bar Y^\rho}|_P{\p Y^\nu \over \p \bar Y^\sigma}|_{P'}\langle \bar Y^\rho (z) \bar Y^\sigma(z')\rangle
  \ee
   (rather than $\langle Y^\mu (z) Y^\nu(z')\rangle$)\footnote{Another way to motivate this is to note that in the ERG the relevant two point function should be thought of as $\langle \delta Y^\mu (z) \delta Y^\nu(z')\rangle$, which is a covariant object}.
 Also note ${\delta S_{int}\over \delta Y^\mu (z)}|_P$ is indeed a tensor (vector) at P because $S_{int}$ is a scalar.

The net effect is that the covariant generalization of the second term (of the ERG) in the flat space ERG can be written schematically as ($\bar Y^\mu$ is the RNC )
\[
\int dz\int dz'~\langle \bar Y^\mu(z) \bar Y^\nu (z')\rangle {\delta S_{int}\over \delta \bar Y^\mu(z)}{\delta S_{int}\over \delta \bar Y^\nu(z')}
\]
\be	\label{ERGRNC}
=
\int dz\int dz'~\langle y^\mu(z) y^\nu (z')\rangle {\delta S_{int}\over \delta  Y^\mu(z)}{\delta S_{int}\over \delta Y^\nu(z')}
\ee

Written in this form it is easy to see that the entire expression is a coordinate scalar.   $S_{int}$ is a scalar and therefore so is the combination
$y^\mu {\delta S_{int}\over \delta Y^\mu}$. 

The first term of the ERG involves second derivatives, but is located at one point and therefore is a local object that can be taken to be located at the origin of the RNC. Thus it is straightforward to evaluate it in the RNC, and then it can be simply covariantized.

\subsection{Covariant OPE} 
There are two ingredients in an OPE: a Taylor expansion and a contraction. 
Thus for example:
\be
e^{ikX(z)}e^{ipX(0)} = e^{ikX(z) + ip X(0)}=e^{ik(X(0)+z\p_z X(0) + \hf z^2 \p_z^2 X(0)+..) +ipX(0)}
\ee
In order to take care of self contractions we can introduce the normal ordered vertex operators by
\be
e^{ikX(z)} = e^{-{k^2\over 2} ln~a}:e^{ikX(z)}:
\ee
and
\[
e^{ikX(z) + ip X(0)}= e^{\hf \langle (ik.X(z)+ipX(0))(ik.X(z)+ipX(0))\rangle} :e^{ikX(z) + ip X(0)}:
\]
\[
= e^{- \hf {(k^2 +p^2)\over 2} ln~a - k.p~ ln~(z^2+a^2)} :e^{ikX(z) + ip X(0)}:
\]
\be
=e^{- \hf {(k^2 +p^2)\over 2} ln~a - k.p~ ln~(z^2+a^2)}:e^{ik(X(0)+z\p_z X(0) + \hf z^2 \p_z^2 X(0)+..) +ipX(0)}:
\ee
We have used a choice of cutoff Green function $G(z,0;a)= \hf ln ~(z^2+a^2)$ for illustration.

\subsubsection{Covariant OPE: Covariant Taylor Expansion and Covariant Contraction}

If one wants a covariant expansion one cannot have contractions between operators at different world sheet locations. Thus to begin with the Green function needs to be Taylor expanded. Thus
$\lan y^\mu (z) y^\nu (0)\ran$ has to be expressed as a power series in $z$ and then each term has to be covariantized.
Thus we write in the RNC (below symmetrization does not have  a normalization factor of $n!$ - so that is explicitly multiplied.)
\[
\bar Y^i (z) = \bar Y ^i (0) + z^\al  \bar Y_\al ^i (0) + {z^\al z^\beta\over 2!} \p_\al  \bar Y^i _\beta(0) + {z^\al z^\beta z^\gamma\over 3!} \p_\al \p_\beta \bar Y^i_\gamma (0)+  {z^\al z^\beta z^\gamma z^\delta \over 4!} \p_\al \p_\beta \p_\gamma  \bar Y^i _\delta(0) +...
\]
\[
= \bar Y ^i (0) + z^\al \bar Y ^i _\al(0) + {z^\al z^\beta\over 2!} D_\al  \bar Y^i _\beta(0) + {z^\al z^\beta z^\gamma\over 3!} D_\al D_\beta  \bar Y^i _\gamma(0)
\]
\be
+  {z^\al z^\beta z^\gamma z^\delta \over 4!}[ D_\al D_\beta D_\gamma \bar Y^i_\delta (0) + {1\over 48} (R^i_{~dac}(0)+ R^i_{~cad}(0))\bar Y^d_{(\delta} (0)\bar Y^c_\gamma (0)\bar D_\beta Y_{\al)}^a(0)]+...
\ee
which is a covariant expansion.  

Thus the Green function is expanded as
\be
G^{ij}(z,0;a)= G^{ij}(0,0;a) + z^\al (\p_\al G^{ij}(z,0;a))|_{z=0}+ {z^\al z^\beta\over 2!}(\p_\al \p_\beta G^{ij}(z,0;a))|_{z=0}+...
\ee
Note that every term is finite because of the presence of a cutoff. Each term involves the metric tensor, Riemann tensor and (covariant) derivatives thereof, all evaluated at one point, which can be taken to be the origin of the RNC.

These Taylor expanded Green functions have to be used for the contractions that are involved in defining OPE of normal ordered vertex
operators as discussed above.

Similar expansions have to be done for the terms in the world sheet action which are products of space time fields and vertex operators.

Before we perform a Taylor expansion we note the following: Since the field strengths are gauge invariant we have no further need of the $\xn$    
and we can set $\xn = 0= \xnb$ and thus $Y^\mu (\xn,z,\zb)= X^(z,\zb)$and $\bar Y^\mu (\xn,z,\zb)\equiv \bar X^(z,\zb)$. Also $\bar Y^\mu_m = {\p_z ^m \bar X^\mu \over (m-1)!}$. In a general coordinate system we can thus set  $Y^\mu_m = {D_z ^m X^\mu \over (m-1)!}$. Thus Taylor expansion will be done in the variable $z$ in the usual manner, except that one has to worry about covariance.

In performing a covariant OPE one also needs to perform a covariant Taylor expansion of the operators.  

Consider first the Taylor expansion of a scalar field about a point O labelled by $X_0$. Let $\Delta X^\mu = X^\mu -X_0^\mu$. Then
\be   \label{ScalarTaylor}
\phi (X) = \phi(X_0) + \Delta X^\mu \p_\mu \phi (X_0) + {1\over 2!} \Delta X^\mu \Delta X^\nu \p_\mu \p _\nu \phi (X_0)+...
\ee

In a general coordinate system $\Delta X^\mu$ is not a tensor, hence this is not a covariant expansion in terms of tensors. The solution is well known: One works in a RNC where $\bar Y^\mu = s \xi ^\mu$. In this case the relation between $X$ and $\bar Y$ is
\be    \label{RNC}
X^\mu = X_0^\mu + \bar Y^\mu -\hf \Gamma^\mu_{\nu \la}|_0 \bar Y^\mu \bar Y^\la - {1\over 3!} \tilde \Gamma ^\mu_{\nu \rho \la}|_0\bar Y^\nu \bar Y^\rho \bar Y^\la ...
\ee
 One can perform a Taylor expansion in powers of $\bar Y^\mu$ to get:
\be   \label{TaylorRNC}
\phi(\bar Y)= \phi (0) + \bar Y^\mu {\p \phi \over \p \bar Y^\mu} +{1\over 2!} \bar Y^\mu \bar Y^\nu {\pp \phi \over \p \bar Y^\mu \p \bar Y^\nu} +...
\ee
Now one can explicitly show that the ordinary derivatives are covariant derivatives because the relevant $\bar \Gamma$ and specific combinations of their derivatives $ \bar {\tilde \Gamma} _{abc..}$,  all vanish. Thus each term is a scalar at the origin O. The LHS is a scalar at P, with coordinates $X$ or $\bar Y$. Furthermore if one transforms this formula to a general coordinate system, then we can think of $\bar Y^\mu$ as a vector at O transformed to the  coordinates, say $X$, and the ordinary derivatives can be replaced by covariant derivatives. Thus one can write
\be   \label{Taylorgeneral}
\phi( Y)= \phi (0) +  Y^\mu \nabla _\mu\phi (0) +{1\over 2!}  Y^\mu  Y^\nu \nabla_\mu \nabla_\nu \phi(0) +...
\ee
where $Y^\mu = {\p \bar Y^\mu \over \p X^\nu}|_0\bar Y^\nu$ is a vector at O.
Again each term is a scalar at the origin. 

In the case of string theory we have $X(z,\bar z)$. Thus when we have $X(z,\bar z)$ and $X(0)$, two different points on the manifold, one can consider a Taylor expansion of a  scalar function in powers of $z^\al$ (rather than $\Delta X$). This is guaranteed to be a covariant expansion in {\em any} coordinate system because $z^\al$ is a coordinate scalar. More explicitly consider the following expansion (we use Latin indices for space time coordinates from now on):
\be   \label{Taylorz}
\phi(X(z))= \phi(X(0)) + z^\al \p_\al \phi(X(0)) + {1\over 2!} z^\al z^\beta \p_\al \p_\beta \phi (X(0))+....
\ee
\[
=\phi(X(0)) + z^\al \p_\al X^i \nabla _i \phi + {1\over 2!} z^\al z^\beta (D_\beta X_\al ^i \nabla_i \phi + X^i_\al X^j_\beta \nabla_i\nabla_j \phi ) + {z^\al z^\beta z^\gamma \over 3!} [{1\over 6} D_{(\al }D_\beta X_{\gamma )}^i] \nabla_i \phi + 
\]
\be  \label{Taylorzz}
{z^\al z^\beta z^\gamma \over 4}
X_\al ^{(i} D_\beta X_\gamma ^{j)} \nabla _i \nabla _j \phi + {z^\al z^\beta z^\gamma \over 3!} X_\al^iX_\beta ^j X_\gamma ^k \nabla _i \nabla_j \nabla_k \phi +...
\ee
Here $D_\al$ is the covariant derivative defined in \eqref{CovDer} and $X^i_\al = \p_\al X^i$.

Note that the equality of \eqref{Taylorz} (which has no $\Gamma$ in it)  and \eqref{Taylorzz} (which superficially has $\Gamma$ dependence being written in terms of covariant derivatives),  implies that there is actually no dependence on $\Gamma$ in\eqref{Taylorzz} - the $\Gamma$ terms cancel amongst themselves. 

We give another example:
\[
S_i (X(z))\p_\al X^i(z) = S_i(X(0)) \p _\al X^i(0) + z^\beta [ S_i {D_{(\beta} X^i_{\al )}\over 2} + \nabla_j S_i X_\al^i X_\beta ^j]+
\]
\be \label{VecTaylorz}
{z^\beta z^\al\over 2!} [ \nabla _jS_i {D_{(\beta} X^i_{\al )}\over 2} X_\al^i + ({\nabla_{(j}\nabla _{k)}S_i\over 2}+{1\over 3} R^l_{jki}S_l)X_\al ^i X_\beta ^j X_\gamma ^k + {D_{(\al} D_\beta X^i_{\gamma )}\over 6}S_i + (\nabla_jS_i){D_{(\beta} X^i_{\al )}}X_\gamma ^j]+...
\ee
 Once again the dependence on $\Gamma$ is completely illusory!
 
 More generally  the vertex operator has higher derivatives and then the starting point involves $D_z^n X(z)$. The expansion now  has a genuine dependence on $\Gamma ^R$. But this dependence is compensated for by the modifications introduced in \eqref{Nontensor1},\eqref{Nontensor} and \eqref{Nontensor1}. However these modifications affect other equations. Thus individual equations are not $\Gamma^R$ independent, although the full theory is.
 
 The upshot is that one can do an ordinary Taylor expansion and simply covariantize all derivatives. The expansions given above are general.
 Since we are using it only for flat backgrounds the curvature tensor can be set to zero.
  
 To summarize, we can then proceed as follows: Evaluate the ERG in the RNC as in the LHS \eqref{ERGRNC}. It involves the scalar $(\bar Y^\mu(z)
 {\delta S_{int}\over \delta Y^\mu(z)})$ at two different points. Thus the final expression can be Taylor expanded about a common point.
 For convenience we can take $\pm z$ as the two points and expand about $z=0$. This Taylor expansion can be covariantized directly by adding $\Gamma ^R$ at appropriate points. This is done by covariantizing derivatives. In addition the massive fields  depend on $\Gamma ^R$ in order to make them tensors. These are as given in \eqref{Nontensor} and \eqref{Nontensor1}. The equations are now manifestly gauge invariant and covariant under BGCT, but  by the general arguments given earlier we know that the full theory does not have any dependence on $h_{\mu\nu}^R$ or $\Gamma^R$.
 
 However this procedure requires that we Taylor expand both the Green function in \eqref{ERG} and the product of operators and the Green function in the contractions due to normal ordering. If we retain the unexpanded Green function, it is hard to see that the full expression is a coordinate scalar.  Of course if we choose to work in a particular coordinate system, RNC,  then one cannot see the manifest coordinate invariance in any case.

\section{Examples}   
We work out some examples below to illustrate the above construction. There are two aspects - the free equation and the interaction term. The interaction term involves a gauge invariant field strength. Once this is computed one an perform an OPE using RNC and covariantize as described in the last section. In this section we compute the free equation and the field strength for various fields in the open and closed string. We also discuss in some detail the field content and dimensional reduction that is necessary in the loop variable approach and outlined as "Step 5" in the Outline. 

\subsection{Open String} As a warm up we reproduce some basic results from I and II on open strings. We also give some comparison with earlier literature on higher spins.

\subsubsection{ Spin 2}

\begin{enumerate}
\item
 For a string in $D$ dimensions, the physical states (defined by light cone oscillators $\al _{-2}^i, \al_{-1}^i\al_{-1}^j$) are $O(D-2)$ tensors given by:
 \ydiagram{1} (4),  ~~\ydiagram{2} (10)
 
 They are combined in the $O(D-1)$ symmetric traceless tensor:
 \ydiagram{2} (15-1=14)
 
 These are the transverse components of a massive $O(D)$ tensor (for which $O(D-1)$ is the little group). $D=26$ for the bosonic string. However for specifying the dimensions of the tensor we use a smaller number for $D$, say $D=6$, which gives manageable numbers. Thus the numbers in brackets are the dimensions of the reps for $D=6$.  
  For a covariant description we keep the trace. For a gauge invariant description more fields are needed - they can be obtained from
 a theory in one higher dimension in the loop variable formalism.
\item {\bf Field Content} (We have used the index "5"  for simplicity to denote the extra dimension.)

\br
\langle \kim \kin \rangle &=& S_{11\mu\nu}\nonumber \\
\langle \kim q_1\rangle &=& S_{11\mu 5} \nonumber \\
\langle q_1 q_1\rangle &=& S_{11}
\er
\item {\bf Gauge Transformation}
\br
\delta S_{11\mu\nu}&=&\langle \delta (\kim \kin)\rangle =\langle \li k_{0(\mu}k_{1\nu )}\rangle  = \p_{(\mu}\Lambda _{11\nu )}
\nonumber \\
\delta S_{11\mu5}&=&\langle \delta (\kim q_1)\rangle = \langle \li \kom q_1 + \li q_0 \kim \rangle = \p_\mu \Lambda _{11} + q_0 \Lambda _{11\mu}\nonumber \\
\delta S_{11} &=&\delta (q_1q_1)= \langle 2 \la _1q_1q_0\rangle = 2\Lambda _{11}q_0 
\er
\item This is exactly what is obtained in earlier literature, for instance \cite{WitFreed}, after a dimensional reduction of massless spin 2, with mass. One can now gauge fix: $S_{11}\to 0$ fixes $\Lambda_{11}$ and $S_{11\mu 5}\to 0$ fixes $\Lambda _{11\mu}$. Thus we are left with $S_{11\mu\nu}$ without any gauge invariance. Thus in terms of irreps this is traceless $S_{11\mu\nu}$ and its trace $S_{11~\mu}^\mu$. This also accords with \cite{SinghHagen} who show that to describe massive spin 2 we need a traceless symmetric 2-tensor and a scalar.

\item In the loop variable approach we also have $\ktm$. 
\br
\lan \ktm \ran &=&S_{2\mu} \nonumber \\
\lan q_2\ran &=& S_2 \\
\er
with gauge transformations:
\br
\delta S_{2\mu} &=& \lan \delta \ktm \ran = \lan \li \kim + \lt \kom \ran = \Lambda  _{11\mu} + \p_\mu \Lambda _2 \nonumber \\
\delta S_2 &=& \lan \delta q_2 \ran = \li q_1 + \lt q_0 = \Lambda _{11}+\Lambda _2
\er
Now in gauge fixing, $\Lambda _{11}$ has been used up. $\Lambda_2$ can be used to set $q_2=0$. That leaves $\ktm$ without any gauge invariance and massive. This is not the right field content of string theory. The resolution is to use the identifications (called $q$-rules in I):
\be
q_1\kim=q_0 \ktm;~~~~\li q_1 = \lt q_0 
\ee
Then $S_{2\mu}$ gets gauged away when we gauge away $S_{11\mu5}$. 

The net effect is that all fields involving $q_n$ can be gauged away. This is consistent with the counting of gauge parameters $\la_n$ that are in 1-1 correspondence with $q_n$.

\item {\bf Gauge Invariant Field Strength:}

Thus in the gauge invariant formulation (after using the q-rules to get rid of $q_1$), we have the fields $\lan \kim \kin\ran = S_{11\mu\nu}, \lan \ktm\ran = S_{2\mu}$, and $\lan q_2\ran = S_2$. As shown in I, for the interacting case we need to introduce $K_{2\mu} = {q_2\over 2 q_0} \kom$ and $K_{11\mu} = \ktm - {q_2\over 2 q_0} \kom$. One can then calculate the gauge invariant field strengths using the prescription given in I.  : $\delta S\over \delta Y^\mu$. This gives for $\ktm$ (dual to $Y_2^\mu$) after simplification:
\br   \label{FS2}
V_{11\mu\nu} &=&\lan \kim\kin - k_{0(\mu}k_{2\nu )} + \underbrace{{q_2\over q_0} \kom}_{K_{2\mu}}\kon \ran\nonumber\\
&=& S_{11\mu\nu} - \p_{(\mu} S_{2\nu)} +\p_\mu\p_\nu {S_2\over q_0}
\er
and dual to $\yim \yin$:
\br   \label{FS3}
V_{11\mu\nu\rho} &=& \lan-\hf \kom \kin \kir +\hf \kim (\kin \kor +\kir \kon) - \underbrace{(\ktm - {q_2\over 2q_0} \kom)}_{K_{11\mu}}\kon\kor\ran \nonumber \\
&=& -\hf \p_\mu S_{11\nu\rho} + \hf \p_{(\rho}S_{11 \nu) \mu} -\p_\nu\p_\rho S_{2\mu} +\p_\mu\p_\nu \p_\rho{S_2\over 2 q_0}
\er
Note that the field strength appears only in the interaction term, so the presence of higher derivatives is expected. Note also the presence of $1\over m$ - which means this makes sense only for massive fields. 

Let us pause to compare with a "field strength" introduced by \cite{WF}:
\be
\Gamma^1 _{ \nu \mu \rho} = \p_\mu \phi_{\nu \rho} - \p_\nu \phi_{\rho \mu} -\p_\rho \phi_{\mu\nu}
\ee
Their {\em massless} field $\phi _{\mu\nu}$ has a gauge transformation $\delta \phi_{\mu\nu} = \p_{(\mu} \xi _{\nu)}$ under which the field strength is {\em not} gauge invariant. Their field strength agrees with \eqref{FS3} if we neglect the $K_{11\mu}$ term.  Similarly if we dimensionally reduce with mass, their field strength agrees with \eqref{FS2} if we neglect the $K_{2\mu}$ term.  

Thus a non zero mass is crucial for constructing an interaction term in the loop variable approach for open strings.

Nevertheless the EOM constructed by \cite{deWF} out of the $\Gamma^1$ is gauge invariant:
\[
W_{\mu \nu}= \p^\rho \p_\rho \phi_{\mu\nu} - \p^\rho \p_\mu \phi_{\rho \nu} - \p^\rho \p_\nu \phi _{\mu \rho} + \p_\mu \p_\nu \phi^\rho_{~\rho}=0
\]

It agrees with the equation of motion obtained using the loop variable approach before dimensional reduction, since they both describe massless spin 2 particles:
\be
\hf k_0^2 \kim\kin - \hf \ko .\ki k_{1(\mu} k_{0\nu)} + \hf \kom \kon \ki .\ki=0
\ee
After dimensional reduction:
\be
\hf (k_0^2 + q_0^2) \kim\kin - \hf \ko .\ki k_{1(\mu} k_{0\nu)} + \hf \kom \kon \ki .\ki -\hf q_0^2 k_{2(\mu}k_{0\nu )} + \hf \kom \kon q_2q_0=0
\ee

The quadratic interaction terms in the equation of motion  involves an OPE of the various field strengths and is manifestly gauge invariant. They can be made covariant under BGCT by covariantizing derivatives in the case that the curvatures are zero, which is the situation in this paper. More general covariantization is described in II.

\end{enumerate}
 \subsubsection{Spin 3}
 \begin{enumerate}
 \item
 The physical states (light cone oscillators $\al_{-3}^i, \al_{-2}^i\al_{-1}^j,\al_{-1}^i\al_{-1}^j\al_{-1}^k$)  are \ydiagram{1} (4), ~~\ydiagram{1}$\otimes$\ydiagram{1} (= \ydiagram{2} (10)$\oplus$ \ydiagram{1,1}(6)) and \ydiagram{3} (20) for a total of 40 states.

 These can be combined into a symmetric traceless 3-tensor and an antisymmetric 2-tensor of $O(D-1)$:
 \ydiagram{3} (35-5=30) , \ydiagram{1,1} (10)
 
 The trace can be kept for a covariant description of the 3-tensor, and a scalar also is required. 
 \item {\bf Field Content}
 
 These fields are obtained from a loop variable description.
 
 Implementing the q-rules we get:
 \br
 \lan \kim \kin \kir \ran &=& S_{111\mu\nu\rho}\nonumber \\
 \lan k_{2[\mu}k_{1\nu]} \ran &= &A_{21[\mu\nu]} \nonumber \\
 \lan k_{2(\mu} k_{1\nu)} = \kim \kin q_1 \ran &=& S_{21(\mu\nu)}\nonumber \\
 \lan k_{3\mu} q_0^2= \kim q_1^2 = (\kim q_2 + \ktm q_1)q_0 \ran &=& S_{3\mu}q_0^2\nonumber \\
 \lan \kim q_2 \ran &=& S_{12\mu}\nonumber \\
 \lan q_3 q_0^2=q_2 q_1 q_0 = q_1^3 \ran &=& S_3 q_0^2
 \er
 \item {\bf Gauge parameters and transformations:}
 
 \br
 \lan \li q_1 q_1 = (\lt q_1 +\li q_2)q_0  = \lambda _3 q_0^2\ran &=& \Lambda _3 q_0^2\nonumber \\
 \lan \li q_2 - \lt q_1 \ran &=& \Lambda _Aq_0\nonumber \\
 \lan \li q_1 \kim = \hf(\lt \kim +\li \ktm ) q_0 \ran &=& (\Lambda _{12\mu} + \Lambda _{21\mu})q_0 = 2q_0\Lambda _S\nonumber \\
 \lan \hf(\lt \kim -\li \ktm )\ran &=& \Lambda _A\nonumber \\
 \lan \li \kim \kin \ran = \Lambda _{111\mu\nu}
 \er
 \br
 \delta S_3 &=& \Lambda _3\nonumber \\
 \delta S_{3\mu} &=& 2 \Lambda _{S\mu} + \p_\mu \Lambda _3\nonumber \\
 \delta (S_{3\mu} q_0 - S_{12\mu}) &=& \Lambda _{A\mu}q_0 + \p_\mu (\Lambda _{21}-\Lambda _3)\nonumber \\
 \delta S_{\mu\nu} &=& \Lambda _{111\mu\nu} + \p_{(\mu } \Lambda _{S\nu)}\nonumber \\
 \delta A_{\mu\nu}&=& \p_{[\mu}\Lambda _{A\nu]}\nonumber \\
 \delta S_{111\mu\nu\rho} &=& \p_{(\mu }\Lambda_{111\nu\rho)}
 \er
 It is clear that all fields except $S_{111\mu\nu\rho}, A_{\mu\nu}$ can be set to zero using the gauge parameters. Since $\Lambda_{11\mu\nu}$ is traceless, the trace of $S_{21\mu\nu}$ also is physical. Thus we have in terms of irreps, a traceless 3-tensor  $S_{111\mu\nu\rho}$, a vector $S_{111\mu \nu}^{~~~\mu}$ and a scalar $S_{21\mu }^{~\mu}$ as the physical fields for a massive spin 3 (in agreement with the general rule of \cite{HagenSingh}), plus a massive antisymmetric tensor.  This counting can also be achieved in a simple way by setting all fields containing $q_n$ to zero (and any field related by q-rules). This would leave $\kim\kin\kir$ and $k_{2[\mu}k_{1\nu]}$ and also the trace $k_2.k_1$ as physical fields.
 
 \item {\bf Field Strength}
 
 The gauge invariant field strength dual to $\yim \yin Y_1^\rho Y_1^\sigma$ in the loop variable approach is
 \be
 V_{111\mu\nu\rho\la}= {1\over 3!} \kom k_{1\la} \kir k_{1\sigma}-{1\over 3!} \kim k_{1(\la} \kir k_{0\sigma)} +{1\over 3} K_{11\mu} k_{1(\la} \kor k_{0\sigma)} - K_{111\mu} \kor k_{0\la} k_{0\sigma}
 \ee
 The first two terms are the same as the (non gauge invariant) fields strength defined by \cite{deWF}
 \[
 \Gamma^{(1)}_{\mu\la \rho \sigma} = \p_\mu \phi_{\la \rho \sigma} - \p_\la\phi_{\mu\rho \sigma}
 \]
 
 The free gauge invariant equation of motion for the massless field in the loop variable approach is (this is the same as in \cite{deWF})
 \be
 [{1\over 3!}\ko^2 \kim\kin\kir - \hf \ko.\ki \kim\kin \kor + \hf \ki .\ki \kim \kon \kor]\yim\yin Y_1^\rho
 \ee
 On dimensional reduction and using the q-rules we get:
 \be
 [{1\over 3!}(\ko^2+q_0^2) \kim\kin\kir - \hf \ko.\ki \kim\kin \kor - {1\over 12}q_0^2 k_{2(\mu} \kin k_{0\rho)}+ {1\over 3!} \ki .\ki k_{1(\mu} \kon k_{0\rho)}+ {1\over 3!}q_0^2 k_{3(\mu}\kon k_{0\mu)}]\yim\yin Y_1^\rho
 \ee
 \end{enumerate}
\subsection{Closed Strings} 
\subsubsection {Level $(1,\bar 1)$: Spin 2}

This has been discussed earlier. As shown there we get a covariant equation for a graviton in a background metric. This agrees with what one obtains from $R_{\mu\nu}=0$ linearized about a background. The "gauge invariant field strength" is given in \eqref{CovGravFS}.
\subsubsection{Level $(2,\bar 2)$: Spin 4}
\begin{enumerate}
\item {\bf Physical States}
The closed string physical states are direct products of the open string states.
We have seen that for open strings the states at level 2 come from a two index traceless symmetric tensor. But the covariant description requires the trace. Thus  we have the diagram   
  \vspace{1cm}

\begin{center}
 \ydiagram{2} 
 $~~~\otimes~$
 \ydiagram{2}
 ~~~=~~
 \ydiagram{4}
 $~~~\oplus$
 \ydiagram{3,1}
 $~~~~\oplus$
 \ydiagram{2,2}
 \end{center}
 \vspace{1cm}
 
The gauge invariant description requires many other tensor fields. In open strings $q_1$ was not allowed and had to be replaced. The corresponding rule for closed strings is that the number of $q_1$'s and $q_{\bar 1}$'s should be equal. 

\item{\bf Field Content and Gauge Transformation}

\begin{itemize}
\item {\bf Scalars}
The allowed combinations are:
\br
\delta S^{2\bar2}=\delta (q_2 q_{\bar 2}) &=& 2\la _2 q_0 q_{\bar 2} + 2 \la _{\bar 2} q_{\bar 0} q_2 \nonumber \\
\delta S^{11\bar 1\bar1}=\delta (q_1^2 q_{\bar 1}^2) & =& q_0^2 q_{\bar 0} 2 \la _2 q_{\bar 2} + q_{\bar 0}^2 q_0 2 \la _{\bar 2} q_2
\er

Here we have used the q-rules separately for the left and right modes separately:
\be 
q_1^2 = q_2 q_0;~~~\li q_1 = \lt q_0
\ee
$q_0$ and $q_{\bar 0}$ are a priori independent. We can choose some definite relation between them, such as
\be   \label{qb0}
q_{\bar 0} \la_{ n} ....=q_0 \la _{ n} ...;~~~q_{\bar 0} \la _{\bar n} ....= - q_0 \la _{\bar n}....
\ee
where the three dots stand for any combination of loop variables involving $\kn, q_n$.

This gives
\br
\delta (q_2 q_{\bar 2}) &=& q_0(2\la _2  q_{\bar 2} - 2 \la _{\bar 2}  q_2 )\nonumber \\
\delta (q_1^2 q_{\bar 1}^2) & =& q_0^3 (2 \la _2 q_{\bar 2} + 2 \la _{\bar 2} q_2)
\er

Thus we have two scalar fields and two gauge parameters. The scalar fields can thus be gauged away - they are Stuckelberg fields.
We can express the gauge parameter in terms of field variations as follows:
\br    \label{scparam}
\lt q_{\bar 2} &=& {1\over 4 q_0} \delta (q_2q_{\bar 2}+ {q_1^2q_{\bar 1}^2\over q_0^2}) \nonumber \\
\la _{\bar 2} q_{2} &=& {1\over 4 q_0} \delta ( {q_1^2q_{\bar 1}^2\over q_0^2}-q_2q_{\bar 2})
\er
In terms of fields:
\br
\Lambda ^{2\bar 2} &=& {1\over 4 q_0}\delta (S^{2,\bar 2} +{S^{11,\bar 1\bar 1}\over \qo^2})\nonumber \\
\Lambda^{\bar 2 2}&=&{1\over 4 q_0}\delta ({S^{11,\bar 1\bar 1}\over \qo^2}-S^{2,\bar 2} )
\er
This relation continues to hold in curved space-time as well.

\item{\bf Vectors}

\br
\delta S_\mu^{2\bar 2}=\delta (\ktm \qtb)&=& 2 \ltb q_{\bar 0} \ktm + \li \kim \qtb + \kom \lt \qtb \nonumber \\
\delta S_\rho^{\bar 2 2}=\delta (\ktrb \qt) &=& 2 \lt q_0 \ktrb + \lib \kirb \qt + \kor \ltb \qt
\er

We have two vectors and four vector gauge parameters. We can thus set 
\be 
\lan \li \kim \qtb\ran=0
\ee
 without any damage to our ability to gauge  away Stuckelberg fields. Thus we get
\be
2 \ltb q_{\bar 0} \ktm = \delta (\ktm \qtb ) - \kom \lt \qtb = \delta (\ktm \qtb) - {\kom\over 4 q_0} \delta ( \qt \qtb + {\qi^2\qib^2\over q_0^2})
\ee
This gives using \eqref{qb0}, an expression for the gauge parameter in terms of field variations:
\br   \label{vparam}
 \ltb  \ktm &=& -{1\over 2q_0}[\delta (\ktm \qtb ) - \kom \lt \qtb] = -{1\over 2q_0}[\delta (\ktm \qtb) - {\kom\over 4 q_0} \delta ( \qt \qtb + {\qi^2\qib^2\over q_0^2})]\nonumber \\
 \lt  \ktrb &=& {1\over 2q_0}[\delta (\ktrb \qt ) - \kor \lt \qtb] = {1\over 2q_0}[\delta (\ktrb \qt) - {\kor\over 4 q_0} \delta ( {\qi^2\qib^2\over q_0^2}- \qt \qtb )]
\er
In terms of fields:
\br
\Lambda^{\bar2 2}_{~~\mu}&=&-{1\over 2q_0}[\delta S^{2,\bar 2}_{\mu} - {\p_\mu\over 4 q_0}\delta (S^{2,\bar 2}+{S^{11,\bar 1\bar 1}\over \qo^2})]\nonumber \\
\Lambda^{2\bar 2}_{~~\rho}&=&-{1\over 2q_0}[\delta S^{\bar 2,2}_{\rho} - {\p_\rho\over 4 q_0}\delta ({S^{11,\bar 1\bar 1}\over \qo^2}-S^{2,\bar 2} )]
\er

This is also true in curved space-time.

\item{\bf 2-Tensors}
\br
\delta S_{\mu\rho}^{2\bar 2}=\delta(\ktm\ktrb )&=& \li \kim \ktrb + \lib \kirb \ktm + \kom \ltb \ktrb + \kor \ltb \ktm \nonumber \\
\delta S^{1\bar 11\bar 1}_{\mu \rho}=\delta(\qi \qib \kim \kirb)&=& \li \qo \kim \qib \kirb + \lib q_{\bar 0} \kirb \qi \kim + \kom \li \qi \qib \kirb + \kor \lib \qi \qib \kim\nonumber\\
\delta S_{\mu\nu}^{11\bar 2}=\delta (\kim \kin \qtb)&=& 2 \ltb q_{\bar 0} \kim \kin + \underbrace{k_{0(\mu} \li k_{1\nu)} \qtb}_{=0}\nonumber \\
\delta S^{\bar 1\bar 1 2}_{\rho \sigma}=\delta (\qt \kirb \kisb)&=& 2 \lt \qo \kirb\kisb +  \underbrace{k_{0(\rho} \lib k_{1\bar \sigma)} \qt}_{=0}\nonumber
\er

Using q-rules $\qi \kirb = q_{\bar 0} \ktrb$, $\qi \kim = \qo \ktm$ we get
\br
\delta(\qi \qib \kim \kirb)&=& \li \qo q_{\bar 0} \kim  \ktrb + \lib q_{\bar 0}\qo \kirb  \ktm + \qo q_{\bar 0}(\kom \lt  \ktrb + \kor \ltb  \ktm)\nonumber \\
&=& \qo^2[\li  \kim  \ktrb - \lib \kirb  \ktm + (\kom \lt  \ktrb - \kor \ltb  \ktm)]
\er
Using \eqref{vparam} and \eqref{scparam} we can write an expression for the tensor parameters:
\br
\lib \kirb \ktm &=& {1\over 2 \qo^2} \delta ( \qo^2 \ktm \ktrb - \qi \qib \kim \kirb ) + {1\over 2\qo} \kor \delta[ {(\qt \qtb + {\qi^2\qib^2\over \qo^2})\over 4}{\kom\over \qo} - \ktm \qtb]\nonumber \\
\li \kim \ktrb &=& {1\over 2 \qo^2}\delta ( \qo^2 \ktm \ktrb + \qi \qib \kim \kirb )-{1\over 2\qo} \kom  \delta [\ktrb \qt + {\kor\over \qo} {(\qt \qtb - {\qi^2\qib^2\over \qo^2})\over 4}]\nonumber\\
\ltb \kim\kin &=& -{1\over 2 \qo}\delta(\kim\kin\qtb)\nonumber\\
\lt \kimb\kinb&=&{1\over 2 \qo}\delta(\kimb\kinb\qt)
\er
In terms of space time fields the first equation reads:
\be
\Lambda ^{\bar 1\bar 1 2}_{~~\rho\mu}= {1\over 2 \qo^2}\delta (\qo^2 S^{2\bar 2}_{\mu\rho}-S^{1\bar 11\bar 1}_{\mu\rho})+{1\over 2\qo} \p_\rho [{\p_\mu\over 4 \qo}\delta ({S^{11\bar 1\bar 1}\over \qo^2}-S^{2\bar 2} )- \delta S^{2\bar 2}_\mu]
\ee
In curved space-time covariant derivatives should be used.
\item{\bf 3-Tensor}

\be
\delta S_{\mu\nu\rho}^{11\bar 2}=\delta(\kim\kin\ktrb) = \lib \kirb \kim\kin + k_{0(\mu}\li k_{1\nu)} \ktrb + \kor \ltb \kim\kin
\ee

Here also one can express the gauge parameter in terms of variations of Stuckelberg fields:
\[
 \lib \kirb \kim\kin=-\delta(\kim\kin\ktrb)-(\kom[ {1\over 2 \qo^2}\delta ( \qo^2 \ktn \ktrb + \qi \qib \kin \kirb )-{1\over 2\qo}\kon  \delta [\ktrb \qt + {\kor\over \qo} {(\qt \qtb - {\qi^2\qib^2\over \qo^2})\over 4}]] \]
 \be
+\mu\leftrightarrow \nu)
+ {\kor\over 2 \qo}\delta(\kim\kin\qtb)
\ee
\[
\bar \Lambda_{~~\rho \mu \nu}^{\bar 1\bar 111}=-\delta S_{\mu \nu \rho}^{1,1,\bar 2}-(\p_\mu [{1\over \qo^2} \delta(\qo^2 S_{\nu \rho}^{2\bar 2}+ S_{\nu \rho}^{1\bar 11\bar 1})-{1\over 2\qo} \p_\nu  \delta [S_{\rho}^{\bar 22}+ {1\over 4\qo}\p_\rho (S^{2\bar 2}-{S^{11\bar 1\bar 1} \over \qo^2})]] +\mu\leftrightarrow \nu)+
\]
\be    \label{Lam}
{1\over 2\qo} \p_\rho\delta S_{\mu\nu}^{11\bar 1}
\ee 
Note that this expression involves higher derivatives in the Stuckelberg fields. When we apply the technique of Section 8 to construct
a gauge invariant EOM, curvature coupling terms involving higher derivatives of the Stuckelberg field will be present. This should not be a problem because these can be gauged away and the physical field - which is the four index tensor -  has only two derivatives acting on it. Note also that every term has powers of $\qo$ in the denominator. Thus non zero masses are required for this to be defined.
This equation continues to be true in curved space-time with the replacement of covariant derivatives in place of ordinary derivatives.

\end{itemize}

We focus on the four index tensor, which contains all the physical states.  The four index tensor is also interesting because it includes as shown above, tensors with mixed symmetry.

The world sheet action has a term $(\ki . Y_1)^2 (k_{\bar 1} . Y_{\bar 1})^2$ corresponding to the 4-tensor:
\be		\label{S1111}
\lan \kim \kin k_{\bar 1 \rho} k_{\bar 1\sigma}\ran = S_{\mu \nu\rho \sigma}^{11\bar1\bar1}
\ee
We can define tensor irreps by writing (brackets denote symmetrization: $S_{(\mu \sigma)}=S_{\mu \sigma} +S_{\sigma \mu}$) the "resolution of unity":
\be
S_{\mu \nu \rho \sigma} = {1\over 24} \underbrace{SS_{\mu \nu \rho \sigma}}_{ \ydiagram{4}}+{1\over 8}\underbrace{S31_{\mu \nu (\rho \sigma)}}_{\ydiagram{3,1}}+{1\over 12}\underbrace{S22_{\mu \nu \rho \sigma}}_{\ydiagram{2,2}}
\ee

 Hereafter, for simplicity we use $S_{\mu\nu\rho \sigma}$ instead of $S ^{11\bar1\bar1}_{\mu\nu\rho \sigma}$. Its gauge variation is
\br
\delta \lan \kim\kin \kirb \kisb \ran &=& \lan \li \kom \kin \kirb \kisb \ran  + \lan \li \kon \kim \kirb \kisb \ran	+\lan \lib \kor \kim \kin  \kisb \ran  +\lan \lib k_{0\sigma} \kim\kin \kirb  \ran \nonumber \\
\implies \delta S_{\mu \nu \rho \sigma}&=& \p_\mu \Lambda _{~~\nu \rho \sigma}^{11\bar 1\bar 1} +\p_\nu \Lambda _{~~\mu \rho \sigma}^{11\bar 1\bar 1}+\p_\rho \bar \Lambda _{~~\sigma \mu \nu }^{\bar 1\bar 111}+\p_\sigma \bar \Lambda _{~~\rho\mu \nu  }^{\bar 1\bar 111}
\er	  
For the gauge transformation parameter $\Lambda_{ \nu \rho \sigma} = \lan \li \kin \kirb \kisb\ran$  irreps are defined by the resolution of unity which reads as:
\be
\Lambda_{\nu \rho \sigma}={1\over 6}\underbrace{\Lambda S _{\nu \rho \sigma}}_{\ydiagram{3}} -{1\over 3}\underbrace{\Lambda I_{\sigma \rho \nu}}_{\ydiagram{2,1}}
\ee
In terms of these fields and gauge parameters one obtains:

\br
{1\over 24}\delta SS_{i_1i_2i_3i_4}&=&{1\over 12}[\p_{i_1}\Lambda S_{i_3i_4i_2}+\p_{i_2}\Lambda S_{i_3i_4i_1}+\p_{i_3}\Lambda S_{i_4i_1i_2}+\p_{i_4}\Lambda S_{i_3i_1i_2}]\nonumber\\
{1\over 8}\delta S31_{i_1i_2(i_3i_4)}&=&{1\over 12}[\p_{(i_1}\Lambda S_{|i_3i_4|i_2)}-\p_{(i_3}\Lambda S_{|i_1i_2|i_4)}]-{1\over 6}[\p_{(i_1}\Lambda I _{|i_3i_4|i_2)}  -  \p_{(i_3}\Lambda I _{|i_1i_2|i_4)}]   \nonumber \\
{1\over 12}\delta S22_{i_1i_2i_3i_4}&=&-{1\over 6}[\p_{(i_1}\Lambda_{|i_3i_4|i_2)}+\p_{(i_3}\Lambda_{|i_1i_2|i_4)}]
\er
There is an identical complex conjugate equation involving $\bar \Lambda$ which we do not bother to write down.
 \item {\bf Free Equation}
   
The free equation of motion (EOM) can be written as:
\[
   -{1\over 4} \ko^2 (\ki .Y_1)^2 (k_{\bar 1} .Y_{\bar 1})^2 +\hf \ko.\ki (\ko .Y_1)(\ki .Y_1) (\kib .\yib)^2 + \hf \ko.\kib (\ko.\yib)(\kib .\yib)(\ki.\yi)^2 +
\]
\be
   -{1\over 4} \ki.\ki (\ko.\yi)^2(\kib.\yib)^2 -{1\over 4} \kib.\kib (\ko.\yib)^2(\ki.\yi)^2 - \ki.\kib (\ko.\yi)(\ko.\yib)(\ki.\yi)(\kib.\yib)
\ee
  It is gauge invariant under 
\[
  \kim \to \kim + \li \kom;~~~~\kimb \to \kimb + \la_{\bar1} \kom
\] 
  if we use the tracelessness condition on the gauge parameters: 
\[
   \li \ki.\kib \kimb= \li \kib.\kib \kim =0= \lib \ki.\kib \kim = \lib \ki.\ki \kimb
\] 
   
   Using \eqref{S1111} the EOM becomes:
   \[
	   -\pp S^{11\bar1\bar1}_{\mu\nu \rho\sigma} + \p^\la \p_{(\mu}S^{11\bar1\bar1}_{\nu)\la \rho \sigma} + \p^\la \p_{(\sigma} S^{11\bar1\bar1}_{|\mu \nu \la|\rho)}
   \]
   \be   \label{FEOM}
   -\p_\mu \p_\nu S^{11\bar1\bar1\la}_{~~~~~~\la\rho\sigma}- \p_\rho \p_\sigma S_{~~~~\mu\nu ~\la}^{11\bar1\bar1~~\la}-\p_{(\sigma}\p_{(\nu}S_{~~~~\mu)~\la|\rho)}^{11\bar1\bar1~~\la}=0
   \ee

  It turns out that an action can also be written for this free theory:
  \[
S_{free}=  -\hf S^{abcd}\Box S_{abcd} - \p_a S^{aefg}\p^bS_{befg} - \p_a S^{efga}\p^bS_{efgb}
  \]
  \[
  -\p_a\p_b S^{abfg}S^c_{~cfg} -\p_a \p_b S^{fgab}S_{fg~c}^{~~c}-4\p_a\p_b S^{eafb}S_{e~fc}^{~c}
  \]
  \[
  +\hf\big(S^{c~fg}_{~c}\Box S^a_{~afg} +S^{fgc}_{~~~~c}\Box S_{fg~a}^{~~a} +4 S^{cf~g}_{~~c}\Box S^a_{~fag}\Big)
  \]
  \[
  +2\Big(S^{c~de}_{~c}\p_d\p^b S_{b~ae}^{~a} +  S^{dec}_{~~~~c}\p_e\p^a S_{d~ba}^{~b}\Big)
  \]
  \be
  -\hf\Big( S^{c~de}_{~c}\p_e\p^a S^b_{~bad} + S^{dec}_{~~~~c}\p_e\p^a S_{ad~b}^{~~b}\Big)
  \ee
 The EOM obtained from this action are linear combinations of \eqref{FEOM} and its traces. 
 
 The action is of the form $S_aM^{ab}S_b$, where $M$ is a symmetric in its indices. Its gauge variation is therefore
 $S_a M^{ab}\delta S_b$. Since $M^{ab} S_b$ is the EOM, which we know is gauge invariant, it must be true that
 $ \delta (M^{ab} S_b)=M^{ab}\delta S_b=0$. Thus it follows that the action is also gauge invariant.

{\bf Covariantization:}
 Covariantization is simple - replace ordinary derivatives by covariant derivatives.

 \item{\bf Interaction}
 
 We now turn to the field strength. It can be compactly written as:
 \[
i\kom {(iK_{1;0}.Y_1)^2\over 2!}{(i K_{0;\bar 1}.Y_{\bar 1})^2\over 2!}
-iK_{1;0\mu} (i\ko.Y_1)(iK_{1;0} .Y_1){(iK_{0;\bar 1}.Y_{\bar 1})^2\over 2!} \]\[
-iK_{0;\bar 1\mu}(i\ko.Y_{\bar 1})(iK_
{0;\bar 1}.Y_{\bar 1}){(iK_{1;0}.Y_1)^2\over 2!}
+iK_{1;\bar 1\mu} (i\ko.Y_1)(i\ko.Y_{\bar 1} )(iK_{1;0}.Y_1)(iK_{0;\bar 1}.Y_{\bar 1})\]
\[+iK_{1, 1;0\mu} (i\ko.Y_1)^2{(iK_{0;\bar 1}.Y_{\bar 1})^2\over 2!}+iK_{0;\bar1, \bar 1\mu} (i\ko.Y_{\bar 1})^2{(iK_{1;0}.Y_{ 1})^2\over 2!}
\]
\[
-iK_{1,1;\bar 1\mu}(i\ko.Y_{\bar 1})(iK_{0;\bar 1}.Y_{\bar 1})(i\ko.Y_{ 1})^2
-iK_{1;\bar 1,\bar 1\mu}(i\ko.Y_{ 1})(i\ko.Y_{\bar 1})^2(iK_{ 1;0}.Y_{  1})
\]
\be	\label{int4}
+iK_{1,1;\bar 1,\bar 1\mu}(i\ko.Y_{\bar 1})^2(i\ko.Y_{ 1})^2
\ee
In this form it is easy to see that it is gauge invariant. The variables $K_{\mu [n];[\bar m]}$ and their gauge transformations were defined in III \cite{BSERGclosed} and are reproduced below:
\br    \label{K}
K_{1;0\mu}&=&	\kim		\nonumber \\
K_{0;\bar 1\mu}&=&	\kimb		\nonumber \\
K_{1;\bar 1\mu}&=&	\bar y_1\kim + y_1 \kimb - y_1\bar y_1 \kom	\nonumber \\
K_{1,1;\bar 1\mu}&=& \bar y_1 (\ktm - y_2 \kom) + {y_1^2\over 2} \kimb -	{y_1^2\over 2} \bar y_1\kom\nonumber \\
 K_{1;\bar 1,\bar 1\mu}&=&y_1(k_{\bar 2 \mu}-\bar y_2 \kom) +  {\bar y_1^2\over 2} \kim -	{\bar y_1^2\over 2}  y_1\kom	\nonumber \\
 K_{1,1;\bar 1,\bar 1\mu}&=&	{y_1^2\over 2}(k_{\bar 2 \mu}-\bar y_2 \kom)+{\bar y_1^2\over 2}(\ktm - y_2\kom)-{ y_1^2\over 2}{\bar y_1^2\over 2}\kom
 \er

Substituting \eqref{K} in \eqref{int4} one obtains for the gauge invariant field strength tensor:

\[F_{\alpha \mu\nu\rho\sigma}=
-k_{0\sigma } k_{1\mu } k_{1\nu } k_{\alpha  \bar{1}} k_{\rho  \bar{1}}-
k_{0\rho } k_{1\mu } k_{1\nu } k_{\alpha  \bar{1}} k_{\sigma  \bar{1}}+
2 k_{0\mu  }k_{0\nu } k_{2 \alpha } k_{\rho  \bar{1}} k_{\sigma  \bar{1}}-
k_{0\nu } k_{1\alpha } k_{1\mu } k_{\rho  \bar{1}} k_{\sigma  \bar{1}}-
\]\[
k_{0\mu } k_{1\alpha } k_{1\nu } k_{\rho  \bar{1}} k_{\sigma  \bar{1}}+
k_{0\alpha } k_{1\mu } k_{1\nu } k_{\rho  \bar{1}} k_{\sigma  \bar{1}}+
2 k_{0\rho }k_{0\sigma } k_{1\mu } k_{1\nu } k_{\alpha  \bar{2}}+
\frac{k_{0\nu } k_{0\sigma }k_{1\mu } k_{\alpha  \bar{1}} k_{\rho  \bar{1}} q_1}{q_0}\]\[+
\frac{k_{0\mu } k_{0\sigma }k_{1\nu } k_{\alpha  \bar{1}} k_{\rho  \bar{1}} q_1}{q_0}+
\frac{k_{0\rho } k_{1\mu } k_{\alpha  \bar{1}} k_{\sigma  \bar{1}} q_1}{q_0}+
\frac{k_{0\rho }\text{  }k_{0\mu } k_{1\nu } k_{\alpha  \bar{1}} k_{\sigma  \bar{1}} q_1}{q_0}-
\frac{2 k_{0\rho }\text{  }k_{0\nu }k_{0\sigma } k_{1\mu } k_{\alpha  \bar{2}} q_1}{q_0}\]\[-
\frac{2 k_{0\rho }\text{  }k_{0\mu }k_{0\sigma }k_{1\nu } k_{\alpha  \bar{2}} q_1}{q_0}-
\frac{k_{0\mu }\text{  }k_{0\nu }k_{0\sigma } k_{\alpha  \bar{1}} k_{\rho  \bar{1}} q_1^2}{q_0^2}-
\frac{k_{0\mu }\text{  }k_{0\nu }k_{0\rho } k_{\alpha  \bar{1}} k_{\sigma  \bar{1}} q_1^2}{q_0^2}+
\frac{k_{0\mu }\text{  }k_{0\nu }k_{0\alpha } k_{\rho  \bar{1}} k_{\sigma  \bar{1}} q_1^2}{q_0^2}\]\[+
\frac{2 k_{0\mu }\text{  }k_{0\nu }k_{0\rho }k_{0\sigma } k_{\alpha  \bar{2}} q_1^2}{q_0^2}-
\frac{2 k_{0\mu }\text{  }k_{0\nu }k_{0\alpha } k_{\rho  \bar{1}} k_{\sigma  \bar{1}} q_2}{q_0}-
\frac{2 k_{0\mu }\text{  }k_{0\nu }k_{0\sigma } k_{2 \alpha } k_{\rho  \bar{1}} q_{\bar{1}}}{q_{\bar{0}}}+
\frac{k_{0\sigma }\text{  }k_{0\nu } k_{1\alpha } k_{1\mu } k_{\rho  \bar{1}} q_{\bar{1}}}{q_{\bar{0}}}\]\[+
\frac{k_{0\sigma }\text{  }k_{0\mu } k_{1\alpha } k_{1\nu } k_{\rho  \bar{1}} q_{\bar{1}}}{q_{\bar{0}}}-
\frac{2 k_{0\mu }\text{  }k_{0\nu }k_{0\rho } k_{2 \alpha } k_{\sigma  \bar{1}} q_{\bar{1}}}{q_{\bar{0}}}+
\frac{k_{0\rho }\text{  }k_{0\nu } k_{1\alpha } k_{1\mu } k_{\sigma  \bar{1}} q_{\bar{1}}}{q_{\bar{0}}}+
\frac{k_{0\rho }\text{  }k_{0\mu}  k_{1\alpha } k_{1\nu } k_{\sigma  \bar{1}} q_{\bar{1}}}{q_{\bar{0}}}\]\[-
\frac{k_{0\alpha }\text{  }k_{0\nu }k_{0\sigma } k_{1\mu } k_{\rho  \bar{1}} q_1 q_{\bar{1}}}{q_0 q_{\bar{0}}}-
\frac{k_{0\mu }\text{  }k_{0\sigma }k_{0\alpha } k_{1\nu } k_{\rho  \bar{1}} q_1 q_{\bar{1}}}{q_0 q_{\bar{0}}}-
\frac{k_{0\nu }\text{  }k_{0\rho }k_{0\alpha } k_{1\mu } k_{\sigma  \bar{1}} q_1 q_{\bar{1}}}{q_0 q_{\bar{0}}}-
\frac{k_{0\mu }\text{  }k_{0\rho }k_{0\alpha } k_{1\nu } k_{\sigma  \bar{1}} q_1 q_{\bar{1}}}{q_0 q_{\bar{0}}}\]\[+
\frac{2 k_{0\alpha }\text{  }k_{0\mu }k_{0\nu }k_{0\sigma } k_{\rho  \bar{1}} q_2 q_{\bar{1}}}{q_0 q_{\bar{0}}}+
\frac{2 k_{0\alpha }\text{  }k_{0\mu }k_{0\nu }k_{0\rho } k_{\sigma  \bar{1}} q_2 q_{\bar{1}}}{q_0 q_{\bar{0}}}+
\frac{2 k_{0\mu }\text{  }k_{0\nu }k_{0\rho }k_{0\sigma } k_{2 \alpha } q_{\bar{1}}^2}{q_{\bar{0}}^2}-
\frac{k_{0\nu }\text{  }k_{0\rho }k_{0\sigma } k_{1\alpha } k_{1\mu } q_{\bar{1}}^2}{q_{\bar{0}}^2}\]\[-
\frac{k_{0\mu }\text{  }k_{0\rho }k_{0\sigma } k_{1\alpha } k_{1\nu } q_{\bar{1}}^2}{q_{\bar{0}}^2}+
\frac{k_{0\alpha }\text{  }k_{0\rho }k_{0\sigma } k_{1\mu } k_{1\nu } q_{\bar{1}}^2}{q_{\bar{0}}^2}+
\frac{k_{0\alpha }\text{  }k_{0\mu }k_{0\nu }k_{0\rho }k_{0\sigma } q_1^2 q_{\bar{1}}^2}{q_0^2 q_{\bar{0}}^2}-
\frac{2 k_{0\alpha }\text{  }k_{0\mu }k_{0\nu }k_{0\rho }k_{0\sigma } q_2 q_{\bar{1}}^2}{q_0 q_{\bar{0}}^2}
\]
\be \label{FStensor1}
-\frac{2 k_{0\alpha }\text{  }k_{0\rho }k_{0\sigma } k_{1\mu } k_{1\nu } q_{\bar{2}}}{q_{\bar{0}}}+
\frac{2 k_{0\alpha }\text{  }k_{0\nu }k_{0\rho }k_{0\sigma } k_{1\mu } q_1 q_{\bar{2}}}{q_0 q_{\bar{0}}}+
\frac{2 k_{0\alpha }\text{  }k_{0\mu }k_{0\rho }k_{0\sigma } k_{1\nu } q_1 q_{\bar{2}}}{q_0 q_{\bar{0}}}-
\frac{2 k_{0\alpha }\text{  }k_{0\mu }k_{0\nu }k_{0\rho }k_{0\sigma } q_1^2 q_{\bar{2}}}{q_0^2 q_{\bar{0}}}
\ee
 In the above expression one has to use q-rules to get rid of unwanted $q_1$'s in any term where  the numbers of $q_1$'s and $q_{\bar 1}$'s are not equal.
 The relevant q-rules are:
 \br
 \qi^2&=&\qt \qo;~~~\qi \kim = \ktm \qo \nonumber ;~~~\li \qi=\lt \qo\\
 \qib^2&=&\qtb q_{\bar 0};~~~\qib \kirb = \ktrb q_{\bar 0};~~~\lib \qib = \ltb q_{\bar 0}
 \er
 This gives:
 
\[F_{\alpha \mu\nu\rho\sigma}=
-k_{0\sigma } k_{1\mu } k_{1\nu } k_{\alpha  \bar{1}} k_{\rho  \bar{1}}-
k_{0\rho } k_{1\mu } k_{1\nu } k_{\alpha  \bar{1}} k_{\sigma  \bar{1}}+
2 k_{0\mu  }k_{0\nu } k_{2 \alpha } k_{\rho  \bar{1}} k_{\sigma  \bar{1}}-
k_{0\nu } k_{1\alpha } k_{1\mu } k_{\rho  \bar{1}} k_{\sigma  \bar{1}}-
\]\[
k_{0\mu } k_{1\alpha } k_{1\nu } k_{\rho  \bar{1}} k_{\sigma  \bar{1}}+
k_{0\alpha } k_{1\mu } k_{1\nu } k_{\rho  \bar{1}} k_{\sigma  \bar{1}}+
2 k_{0\rho }k_{0\sigma } k_{1\mu } k_{1\nu } k_{\alpha  \bar{2}}+
{k_{0\nu } k_{0\sigma }k_{2\mu } k_{\alpha  \bar{1}} k_{\rho  \bar{1}} }\]\[+
{k_{0\mu } k_{0\sigma }k_{2\nu } k_{\alpha  \bar{1}} k_{\rho  \bar{1}} }+
{k_{0\rho }k_{0\nu} k_{2\mu } k_{\alpha  \bar{1}} k_{\sigma  \bar{1}} }+
{k_{0\rho }\text{  }k_{0\mu } k_{2\nu } k_{\alpha  \bar{1}} k_{\sigma  \bar{1}} }-
{2 k_{0\rho }\text{  }k_{0\nu }k_{0\sigma } k_{2\mu } k_{\alpha  \bar{2}} }\]\[-
{2 k_{0\rho }\text{  }k_{0\mu }k_{0\sigma }k_{2\nu } k_{\alpha  \bar{2}} }-
\frac{k_{0\mu }\text{  }k_{0\nu }k_{0\sigma } k_{\alpha  \bar{1}} k_{\rho  \bar{1}} q_2}{q_0}-
\frac{k_{0\mu }\text{  }k_{0\nu }k_{0\rho } k_{\alpha  \bar{1}} k_{\sigma  \bar{1}} q_2}{q_0}+
\frac{k_{0\mu }\text{  }k_{0\nu }k_{0\alpha } k_{\rho  \bar{1}} k_{\sigma  \bar{1}} q_2}{q_0}\]\[+
\frac{2 k_{0\mu }\text{  }k_{0\nu }k_{0\rho }k_{0\sigma } k_{\alpha  \bar{2}} q_2}{q_0}-
\frac{2 k_{0\mu }\text{  }k_{0\nu }k_{0\alpha } k_{\rho  \bar{1}} k_{\sigma  \bar{1}} q_2}{q_0}-
{2 k_{0\mu }\text{  }k_{0\nu }k_{0\sigma } k_{2 \alpha } k_{\rho  \bar{2}}} +
{k_{0\sigma }\text{  }k_{0\nu } k_{1\alpha } k_{1\mu } k_{\rho  \bar{2}} }\]\[+
{k_{0\sigma }\text{  }k_{0\mu } k_{1\alpha } k_{1\nu } k_{\rho  \bar{2}} }-
{2 k_{0\mu }\text{  }k_{0\nu }k_{0\rho } k_{2 \alpha } k_{\sigma  \bar{2}} }+
{k_{0\rho }\text{  }k_{0\nu } k_{1\alpha } k_{1\mu } k_{\sigma  \bar{2}} }+
{k_{0\rho }\text{  }k_{0\mu}  k_{1\alpha } k_{1\nu } k_{\sigma  \bar{2}} }\]\[-
\frac{k_{0\alpha }\text{  }k_{0\nu }k_{0\sigma } k_{1\mu } k_{\rho  \bar{1}} q_1 q_{\bar{1}}}{q_0 q_{\bar{0}}}-
\frac{k_{0\mu }\text{  }k_{0\sigma }k_{0\alpha } k_{1\nu } k_{\rho  \bar{1}} q_1 q_{\bar{1}}}{q_0 q_{\bar{0}}}-
\frac{k_{0\nu }\text{  }k_{0\rho }k_{0\alpha } k_{1\mu } k_{\sigma  \bar{1}} q_1 q_{\bar{1}}}{q_0 q_{\bar{0}}}-
\frac{k_{0\mu }\text{  }k_{0\rho }k_{0\alpha } k_{1\nu } k_{\sigma  \bar{1}} q_1 q_{\bar{1}}}{q_0 q_{\bar{0}}}\]\[+
\frac{2 k_{0\alpha }\text{  }k_{0\mu }k_{0\nu }k_{0\sigma } k_{\rho  \bar{2}} q_2 }{q_0 }+
\frac{2 k_{0\alpha }\text{  }k_{0\mu }k_{0\nu }k_{0\rho } k_{\sigma  \bar{2}} q_2 }{q_0 }+
\frac{2 k_{0\mu }\text{  }k_{0\nu }k_{0\rho }k_{0\sigma } k_{2 \alpha } q_{\bar{2}}}{q_{\bar{0}}}-
\frac{k_{0\nu }\text{  }k_{0\rho }k_{0\sigma } k_{1\alpha } k_{1\mu } q_{\bar{2}}}{q_{\bar{0}}}\]\[-
\frac{k_{0\mu }\text{  }k_{0\rho }k_{0\sigma } k_{1\alpha } k_{1\nu } q_{\bar{2}}}{q_{\bar{0}}}+
\frac{k_{0\alpha }\text{  }k_{0\rho }k_{0\sigma } k_{1\mu } k_{1\nu } q_{\bar{2}}}{q_{\bar{0}}}+
\frac{k_{0\alpha }\text{  }k_{0\mu }k_{0\nu }k_{0\rho }k_{0\sigma } q_1^2 q_{\bar{1}}^2}{q_0^2 q_{\bar{0}}^2}-
\frac{2 k_{0\alpha }\text{  }k_{0\mu }k_{0\nu }k_{0\rho }k_{0\sigma } q_2 q_{\bar{2}}}{q_0 q_{\bar{0}}}
\]
\be \label{FStensor}
-\frac{2 k_{0\alpha }\text{  }k_{0\rho }k_{0\sigma } k_{1\mu } k_{1\nu } q_{\bar{2}}}{q_{\bar{0}}}+
\frac{2 k_{0\alpha }\text{  }k_{0\nu }k_{0\rho }k_{0\sigma } k_{2\mu }  q_{\bar{2}}}{ q_{\bar{0}}}+
\frac{2 k_{0\alpha }\text{  }k_{0\mu }k_{0\rho }k_{0\sigma } k_{2\nu }  q_{\bar{2}}}{q_{\bar{0}}}-
\frac{2 k_{0\alpha }\text{  }k_{0\mu }k_{0\nu }k_{0\rho }k_{0\sigma } q_2 q_{\bar{2}}}{q_0 q_{\bar{0}}}
\ee
 
 In terms of fields this becomes:
 \[F_{\al \mu \nu \rho \sigma}= -\p_\sigma S^{11\bar 1 \bar 1}_{\mu\nu\al \rho}- S_{\mu \nu \al \sigma}^{11\bar 1 \bar 1}+2 \p_\mu\p_\nu S_{\al \rho \sigma}^{2\bar 1 \bar 1}-S_{\nu \al \mu \rho}^{11\bar 1 \bar 1} \]
 \[-\p_\mu S_{ \al \nu \rho\sigma}^{11\bar 1 \bar 1}+ \p_\al S_{\mu \nu  \rho\sigma}^{11\bar 1 \bar 1}+2 \p_\rho\p_\sigma S_{\mu\nu \al}^{11\bar 2}+ \p_\nu \p_\sigma S_{\mu \al \rho}^{2\bar 1\bar1}
 \]
\[+ \p_\mu \p_\sigma S^{2 \bar 1\bar 1}_{\mu \al \sigma}+\p_\rho \p_\nu S^{2 \bar 1\bar 1}_{\mu \al \sigma}+ \p_\rho \p_\mu S^{2 \bar 1\bar 1}_{\nu \al \sigma}-2 \p_\rho\p_\nu \p_\sigma S^{2\bar 2}_{\mu \al}
\]
\[-2 \p_\rho \p_\mu \p_\sigma S^{2\bar 2}_{\nu \al}- {1\over \qo}\p_\mu\p_\nu\p_\sigma S^{2\bar 1\bar1}_{~\al \rho}-{1\over \qo}\p_\mu\p_\nu\p_\rho S^{2\bar 1\bar1}_{~\al \sigma} +{1\over \qo}\p_\mu\p_\nu\p_\al S^{2\bar 1\bar1}_{~ \rho\sigma}
\]  
 \[+{2\over \qo} \p_\mu\p_\nu\p_\rho\p_\sigma S^{2\bar 2}_{~\al}-{2\over \qo} \p_\mu\p_\nu\p_\al S^{2\bar 1\bar1}_{~\rho\sigma} -2 \p_\nu \p_\mu \p_\sigma S^{2\bar 2}_{ \al\rho}+ \p_\nu\p_\sigma S_{\al \mu\rho}^{11\bar 2}
 \]
 \[ +\p_\mu\p_\sigma S_{\al \nu\rho}^{11\bar 2}-2 \p_\nu \p_\mu \p_\rho S^{2\bar 2}_{ \al\sigma}+\p_\nu\p_\rho S_{\al \mu\sigma}^{11\bar 2}+\p_\mu\p_\rho S_{\al \nu\sigma}^{11\bar 2}
 \]\[-{1\over \qo q_{\bar 0}}\p_\al\p_\nu\p_\sigma S^{11\bar 1\bar1}_{~\mu\rho}-{1\over \qo q_{\bar 0}}\p_\al\p_\mu\p_\sigma S^{11\bar 1\bar1}_{~\nu\rho}-{1\over \qo q_{\bar 0}}\p_\al\p_\nu\p_\rho S^{11\bar 1\bar1}_{~\mu\sigma}-{1\over \qo q_{\bar 0}}\p_\al\p_\mu\p_\rho S^{11\bar 1\bar1}_{~\nu\sigma}
 \]
 \[+{2\over \qo} \p_\al \p_\mu\p_\nu\p_\sigma S^{2\bar 2}_{~\rho}+{2\over \qo} \p_\al \p_\mu\p_\nu\p_\rho S^{2\bar 2}_{~\sigma}+{2\over q_{\bar 0}}  \p_\mu\p_\nu\p_\rho \p_\sigma S^{2\bar 2}_{\al}- {1\over q_{\bar 0}} \p_\nu\p_\rho\p_\sigma S^{11\bar 2}_{\al \mu}
 \]
 \[- {1\over q_{\bar 0}} \p_\mu\p_\rho\p_\sigma S^{11\bar 2}_{\al \nu}- {1\over q_{\bar 0}} \p_\al\p_\rho\p_\sigma S^{11\bar 2}_{ \mu\nu}+{1\over \qo^2 q_{\bar 0}^2}\p_\al
 \p_\mu\p_\nu\p_\rho\p_\sigma S^{11\bar 1\bar 1}-{4\over \qo q_{\bar 0}}\p_\al
 \p_\mu\p_\nu\p_\rho\p_\sigma S^{2\bar 2}
 \]\[ +{2\over q_{\bar0}} \p_\al \p_\nu\p_\rho\p_\sigma S^{2\bar 2}_{\mu}+{2\over q_{\bar0}} \p_\al \p_\mu\p_\rho\p_\sigma S^{2\bar 2}_{\nu}
 \]

 The expression is symmetric in $\mu \leftrightarrow \nu$ and also $\rho \leftrightarrow \sigma$. It is also symmetric under interchange of barred and unbarred variables.

 \item{\bf Example of Contribution to EOM:}
 
 Finally one can input all the above ingredients and write in the RNC:
 \be
\int dz~ \dot G^{\al \beta}(z,0;a)F_{\al \mu_1 \nu_1 \rho_1 \sigma_1}(Y(z))\bar Y_1^{\mu_1}\bar Y_1^{\nu_1}   \bar Y_{\bar 1}^{\rho_1} \bar Y_{ \bar 1} ^{\sigma_1}(z) F_{\beta \mu_2 \nu_2 \rho_2 \sigma_2 }(Y(0))\bar Y_1^{\mu_2}\bar Y_1^{\nu_2}   \bar Y_{\bar 1}^{\rho_2} \bar Y_{ \bar 1}^{ \sigma_2}(0)
 \ee 
Here $F$ is the gauge invariant field strength tensor given in \eqref{FStensor}.
 This will, on Taylor expanding and contracting, produce a lot of terms. They can be organized in terms of normal ordered vertex operators. Thus for example, contributing to the graviton equation, proportional to vertex operator $\e  Y_1^\mu  Y_{\bar 1}^\rho(0)$,  one finds contributions from the above term, such as:
 \be   \label{eom}
 \dot G^{\al \beta}(0,0;a)F_{\al \mu_1 \nu_1 \rho_1 \sigma_1}(Y(0))F_{\beta \mu_2 \nu_2 \rho_2 \sigma_2 }(Y(0))\bar Y_1^{\mu_1} \bar Y_{\bar 1}^{\rho_2} \lan \bar Y_1^{\nu_1}   \bar Y_{\bar 1}^{\rho_1} \bar Y_{ \bar 1} ^{\sigma_1}(0) \bar Y_1^{\mu_2}\bar Y_1^{\nu_2}   \bar Y_{ \bar 1}^{ \sigma_2}(0)\ran 
 \ee
 In the above expression, $F$ can be covariantized using the techniques given in the preceding subsection (and illustrated with two examples). The contractions are finite because a cutoff Green function is used. \eqref{eom} is thus a gauge invariant and {\em background covariant} contribution to the graviton ($\tilde h_{\mu \nu}$) equation of motion. It can be thought of as a contribution to $T_{\mu\nu}$, the RHS of Einstein's equation. The LHS was given in \eqref{GravFreeCov1}. In this approach the graviton is treated just as another string mode and the equations are quadratic in the graviton. GCT is expected to be realized only in the continuum limit.

 The  EOM are non universal, because they depend on the regulator. This ambiguity corresponds to the field redefinitions in the space time field theory. The universality of the continuum limit
of the world sheet theory corresponds, in the space time field theory, to the statement that the on-shell S-matrix is unaffected by field redefinitions.

 \end{enumerate}   
\section{Conclusions}

A technique for writing down background covariant and gauge invariant equation of motion for all the fields of the closed string has been described. This is based on the exact renormalization group (ERG)  applied to a completely general world sheet action. Loop variable techniques are used to make it gauge invariant for all the massive modes. The basic idea is to write a regulated world sheet theory that has 
general coordinate invariance under  transformations that also involve a background metric. One has also to ensure that the full action does not depend on this metric. One can then work in RNC, obtain gauge invariant equations and then covariantize. The background metric has to be diffeomorphic to the Minkowski metric i.e. curvature has to be zero. Thus this technique, while it gives equations with all the necessary local symmetries, is not background independent. The equations for all the modes including the graviton are quadratic.

The gauge transformations in this approach start off being the same for the free and interacting theory. The interactions are written in terms of gauge invariant field strengths. The theory looks "Abelian". But unlike in the case of open strings, here one finds that if the graviton is to be described by a massless equation a gauge invariant field strength cannot be written down and it  is necessary to include general coordinate transformations in the definition of the gauge transformation.  Thus the interacting theory is forced to have general coordinate invariance. One obtains a covariant (quadratic) equation for a graviton fluctuating about a flat background metric. The free equation is exactly what is obtained by linearizing Einstein's equation about a background metric. 

The world sheet theory also has all the massive modes turned on. It can be written in a coordinate invariant way by introducing a background metric. The dependence on the background metric can be removed completely by absorbing them into the definition of the massive fields. Thus the original massive fields have non tensorial transformation laws, but the final RG equations involve redefined fields and have manifest background general covariance.
 The solutions to this equation give the fixed points of the world sheet theory. It is expected that the fixed points, being physically significant, cannot depend on the arbitrary background metric because the action does not. These equations should therefore be generally covariant.

This was  illustrated in some detail for the massive level (2,2) physical field of the closed string where there are fields of mixed symmetry.
In flat space a (free) action was also written down for these fields.

 Thus to summarize: we obtain gauge and (background) covariant equations for all the modes (massive and massless) of the closed string. The equations are written as fluctuations about a flat background and are quadratic in all fields including the graviton.

There are some interesting open questions:

\begin{itemize}
\item

We were restricted to perturbations about flat backgrounds for the following reason: The world sheet action that is used for the interacting theory
is written in a different form using the identity $\p_{\xn} \p_{\xm} Y^\mu = \p_{x_{n+m}}Y^\mu$. This turns out to be necessary for gauge invariance of the field strength. This was described in I,II,III. In curved space the covariant versions of these do not obey this identity.
Thus there is a clash between general covariance and gauge invariance. This needs to be resolved if one wants a background independent formalism.

\item
The linearized gauge transformation of the massless mode got linked to coordinate transformations when we attempted to write an interacting theory. (In III it was also speculated that space time should be complex for a fuller interpretation.) It is tempting to speculate that similar interpretations await the massive gauge transformations. In the present construction, they are realized as spontaneously broken symmetries involving Stuckelberg fields. There may be an extension to this, where it is realized linearly, and would possible involve the extra coordinate $q$ in a more geometric way. It is noteworthy that at the free level the theory does look like a massless higher dimensional theory. This may also cast some light on the original speculation regarding the underlying symmetries made in \cite{BSLV}.

\item   
Finally this construction only gives equations of motion. It remains to be seen whether something on the lines of \cite{BSPA}  can be done to obtain an action.  
   \end{itemize}
 
\end{document}